\documentclass[journal,twoside,final]{IEEEtran}
\usepackage{amsthm}
\usepackage{colortbl}
\usepackage[usenames,dvipsnames]{xcolor}
\usepackage{bbm}
\usepackage{cite}
\usepackage{amssymb,amsmath}
\usepackage{times}
\usepackage{latexsym}
\usepackage{graphicx}
\usepackage{flushend}
\usepackage{bm}
\usepackage{stfloats}
\usepackage{cases}
\usepackage{array}
\usepackage{setspace}
\usepackage{fancyhdr}
\usepackage[small]{caption}
\usepackage{soul}
\usepackage{subfig}
\usepackage{fixltx2e}
\usepackage{multicol}
\usepackage{latexsym}

\allowdisplaybreaks

\begin{document}

\title{A Multi-Channel Spectrum Sensing Fusion Mechanism for Cognitive Radio Networks: \\Design and Application to IEEE 802.22 WRANs}
\author{Navid~Tadayon,~\IEEEmembership{Student Member,~IEEE}, and Sonia~A\"issa,~\IEEEmembership{Senior Member,~IEEE}
\thanks{Manuscript received May 29 2015; revised September 2015; accepted January 29, 2016. This work was supported by a Discovery Grant from the Natural Sciences and Engineering Research Council (NSERC) of Canada.}
\thanks{The authors are with the Institut National de la Recherche Scientifique (INRS), University of Quebec, Montreal, QC, Canada (email: \{tadayon, aissa\}@emt.inrs.ca).}
}

\markboth{IEEE Transactions on Cognitive Communications and Networking, accepted for publication, January 2016.} {Tadayon \MakeLowercase{and} Aissa: A Multi-Channel Spectrum Sensing Fusion Mechanism for Cognitive Radio Networks}
\maketitle

\pubidadjcol

\begin{abstract}
\noindent The IEEE 802.22 is a new cognitive radio standard that is aimed at extending wireless outreach to  rural areas. Known as wireless regional area networks, and designed based on the not-to-interfere spectrum sharing model, WRANs are channelized and centrally-controlled networks working on the under-utilized UHF/VHF TV bands to establish communication with remote users, so-called customer premises equipment (CPEs). Despite the importance of reliable and interference-free operation in these frequencies, spectrum sensing fusion mechanisms suggested in IEEE 802.22 are rudimentary and fail to satisfy the stringent mandated sensing requirements. Other deep-rooted shortcomings are performance non-uniformity over different signal-to-noise-ratio regimes, unbalanced performance, instability and lack of flexibility. Inspired by these observations, in this paper we propose a distributed spectrum sensing technique for WRANs, named multi-channel  learning-based distributed sensing fusion mechanism (MC-LDS). MC-LDS is demonstrated to be self-trained, stable and to compensate for fault reports through its inherent reward-penalty approach. Moreover, MC-LDS exhibits a better uniform performance in all traffic regimes, is fair (reduces the false-alarm/misdetection gap), adjustable (works with several degrees of freedom) and bandwidth efficient (opens transmission opportunities for more CPEs). Simulation results and comparisons unanimously corroborate that MC-LDS outperforms IEEE 802.22 recommended algorithms, i.e., the AND, OR and VOTING rules.
\end{abstract}

\section{Introduction}
After a decade of painstaking and enduring efforts, cognitive radio (CR) is now mature enough to be exploited as the solution to spectrum inefficiency. The spread and diversity of investigations have been substantial in both academia and standardization bodies. The consolidation of these efforts has recently appeared in the form of a prominent standard, IEEE 802.22 for WRANs, which aims at diminishing the rural-urban divide by extending the reach of technology as far as possible. At this point, CR seems to be just a tiny step away from commercialization and transformation into something beyond a technological concept. For this very reason, researchers in both academia and industry need to designate particular attention to  IEEE 802.22 standard by refining its features and capabilities, and introducing compatible, yet more efficient, methods and mechanisms. To incite curious minds, the said needs have been explicitly stated in the form of open issues in this standard. Among these, the distributed sensing fusion mechanism (SFM) is fundamental.\footnote{Hereafter, terms ``sensing-data fusion", ``decision combining" and ``distributed sensing" are used interchangeably as they represent the same concept.}

The standard proposals for the SFM are basically AND, OR and VOTING logic rules. Given the significance of the spectrum sensing process for networks operating in VHF/UHF bands, these combining rules are not satisfactorily reliable, accurate and stable. In particular, they suffer from unbalanced performance, where a low false-alarm rate comes at the cost of an unbearably high misdetection rate (and vice versa), and exhibit rigid performance, that is, they are incapable of adjusting their performance to  varying conditions. Moreover, due to the lack of performance stabilizing feedback from sensing output to input, operational instability is possible, manifesting itself in the form of weak performance in low signal-to-noise ratio (${\rm SNR}$) regimes. Therefore, the need for an efficient distributed SFM that avoids the above shortcomings, complies with the standard directives, and can be easily embedded in the system, is entirely felt.
\begin{table*}[t!]
\begin{center}
\begin{small}
\caption{Main acronyms and symbols in the paper.}
\begin{tabular}{p{50pt}p{110pt}|p{50pt}p{110pt}}
\hline \hline
Acronym & Expanded Form & Acronym & Expanded Form \\
\hline
CH		   & channel           & DCL 	& disallowed CL\\
CL		   & CH list           & LPS	& local priority set \\
${\rm QP}$ & quiet period      &  RB	& resource block \\
OCL		   & operating CL      & CBP	& coexistence beacon protocol   \\
PCL		   & protected CL      & SCH	& superframe control header \\
BCL		   & backup CL         &  CCL   & candidate CL   \\
\hline \hline
Symbol & Definition & Symbol & Definition \\
\hline
$_{i}$, $_{j}$, $_{k}$ & CPE, WRAN, CH indices & $P_{{\rm MD},i,j}^{(t),k}$ & misdetection prob. \\
$\mathfrak{M}$  & $\#$ cells & $P_{{\rm FA},i,j}^{(t),k}$ & false-alarm prob.\\
$\mathfrak{B}$ & $\#$ CHs & $d^{(t)}_{i,j,k}$ & local sensing decision \\
 $\Theta_{i}$ & logistic estimator parameter & ${\chi_{j,k}^{(n)}}^2$ &  Pearson's test stat. \\
$^{(t)}$ & ${\rm QP}$ index & $m_{j,k}$ & $\#$ sensors in ${\rm WRAN}_{j}$ \\
$\mathcal{H}_{1,k}^{(t)}$ ($\mathcal{H}_{0,k}^{(t)}$) & alternative (null) hypotheses & $\mathcal{N}_i$  & $\#$ training samples \\
$\mathcal{S}_{i,j,k}^{(t)}$ & received power & $\beta^{(t)^{\rm Rep}}_{i,j,k}$ & reporting CH coefficient \\
$\lambda_{j,k}$ & power threshold &  $\mathcal{M}_i$ & $\#$ signal samples \\
${\rm IAR}$		& incumbent activity rate & $\mathcal{S}_N$  & noise power \\
${\rm IAF}$ 		& incumbent alteration freq. & $R^{(t)}_{j,k}$  & database reading \\
${L}_{i,j,k}^{(t)}$	& reward-penalty score & $D^{(t)}_{j,k}$  & central decision \\
$\alpha_{j,k}$	& temporal discount & $N_{j,k}$ & historic count  \\
$w_{i,j,k}^{(n)}$	& confidence metric & $X_{i,j,k}^{(n)}$ & CH indicator \\
$\mathcal{Z}_{j}^{(t)}$	& factual CH status\\ \hline
\end{tabular}
\label{tab0}
\end{small}
\end{center}
\end{table*}

The SFM proposed in this paper is named {\it multi-channel learning-based distributed sensing (MC-LDS)} for its dependence on learning through observation fostered by a reward-penalty rationality. For the latter feature, MC-LDS showcases a stable and self-trained behavior as well as robustness in detecting faulty sensing reports. The other salient characteristics of MC-LDS are improved performance  in all traffic regimes, fairness (reduced false-alarm/misdetection gap), adjustability (operability with several degrees of freedom) and  bandwidth efficiency (increased transmission opportunities for more CPEs. Simulation results and comparisons prove that MC-LDS surpasses the sensing schemes proposed in the IEEE 802.22 standard. Furthermore, for its generic architecture, MC-LDS can be integrated to boost the sensing performance of other promising technologies and standards such as \textit{White-Fi} (IEEE 802.11af WLANs), wireless personal area networks and ZigBee (IEEE 802.15 family), cognitive WiMax (IEEE 802.16h) and the recent IEEE 1900.6b standard emerged to support spectrum databases using sensing information.

Before detailing the proposed multi-channel learning-based distributed sensing mechanism, Section II describes the salient proposals on collaborative spectrum sensing, in general, and those for IEEE 802.22 WRANs. The discussion is followed by describing the spectrum sensing function (SSF) of IEEE 802.22 and its related functionality.  MC-LDS is presented in Section III. Validation results and discussions are provided in Section IV, before concluding the paper in Section IV.
\section{Spectrum Management in IEEE 802.22}
When the sensing outcome is purely reliant on local decisions of individual sensors, cost is inexorably a concern. To have an accurate sensing outcome, each sensor node needs to be equipped with a wideband antenna and an amplifier with a wide dynamic range. On the baseband side, high resolution analog-to-digital converters operating over wide frequency bands and ultra-fast signal processing units (DSPs and FPGAs) are essential. Such elements are expensive and have limitations. Performance-wise, local sensing is not reliable and needs many samples, particularly, in the presence of deteriorating dynamic phenomena such as multipath fading, shadowing, noise, and hidden primary users (PUs). To tackle these issues, the broadcasting nature of wireless channels and the spatial diversity can be exploited to distribute the heavy sensing task among sensor nodes across the network. Compared to local sensing, collaborative sensing results in less frequent false-alarm events for a fixed probability of misdetection (and vice versa) and reduces the average detection time.

In this section, some of the most recent investigations on cooperative spectrum sensing in generic settings as well as for WRANs are discussed, followed by the most relevant WRAN's spectrum management and scheduling functionality whose knowledge is essential for understanding the proposed MC-LDS technique elaborated in Section III.
\subsection{Literature Review}
Spectrum aimed for the operation of WRANs \cite{IEEE80222} hugely overlaps with UHF/VHF TV bands in the 54-862 MHz range \cite{Kolodzy2003}. Though these bands have been exclusively conceded to incumbent operators for interference-free operation, they are often empty and under-utilized, in time and space. Yet, the exclusive access right legally inhibits others to use them. With the new insights about what has provoked such spectrum inefficiency, now WRAN operators are authorized to transmit on these bands as long as they do not cause harmful interference to the incumbent service. Such paradigm relies on reliable and accurate channel sensing prior to transmission. To enhance reliability and accuracy, collaborative spectrum sensing has been proposed and shown to alleviate the destructive effects of the hidden/exposed-terminal problems and the channel fading (see e.g. \cite{Sahai2006,Hamza2014} and references therein).

While collaborative spectrum sensing improves the sensing accuracy, this comes at the cost of lower sensing efficiency triggered by higher energy consumption and signalling overhead \cite{Zhang2012}.  Although, high energy consumption may not be an issue in ordinary networks, low power consumption and network livability are the main objectives when nodes are battery powered (e.g. cognitive sensor networks). Stimulated by this idea, \cite{Maleki2011} introduced a combined sleeping/censoring scheme to reduce sensing energy consumption. When the cognitive radio network (CRN) operates over multiple channels, an important issue is to determine how to distribute a limited number of sensors to sense those channels. Two recent studies addressing this problem are \cite{Eryigit2013} and \cite{Li2014} where the first aims at improving energy efficiency and the latter adopts a Bayesian decision rule to maximize the system throughput. Readers are referred to \cite{Yucek2009} for a comprehensive review on sensing for CRNs.
\begin{table*}[t]
\begin{center} \small
\caption{A complete dynamic channel list (CL) for a network of 12 coexisting WRANs, extracted at a random time from the simulator. $\{ \}$ indicates an empty list.}
\begin{tabular}{>{\columncolor[gray]{0.8}} p{30pt} c c c c c c | p{50pt}p{50pt}p{50pt}p{50pt}p{50pt}}
\rowcolor[gray]{0.8}
\cellcolor{white}
&
Operating CL &
Disallowed CL &
Backup CL &
Protected CL &
Candidate CL \\
\hline
WRAN$_{1}$&
{$\left\{ {\rm CH}_{4} \right\}$}&
{$\left\{\right\}$}&
{$\left\{\right\}$}&
{$\left\{ {\rm CH}_{2}, {\rm CH}_{10} \right\}$}&
$\left\{\right\}$
\\
WRAN$_{2}$&
{$\left\{ {\rm CH}_{8} \right\}$}&
{$\left\{ {\rm CH}_{10} \right\}$}&
{$\left\{\rm  {\rm CH}_{4} , {\rm CH}_{6} \right\}$}&
{$\left\{\right\}$}&
${\left\{\right\}}$
\\
WRAN$_{3}$&
{$\left\{ {\rm CH}_{4} \right\}$}&
{$\left\{\right\}$}&
{$\left\{\right\}$}&
{$\left\{ {\rm CH}_{2} \right\}$}&
${\left\{ {\rm CH}_{5} \right\}}$
\\
WRAN$_{4}$&
{$\left\{ {\rm CH}_{10} \right\}$}&
{$\left\{ {\rm CH}_{10} \right\}$}&
{$\left\{\right\}$}&
{$\left\{\right\}$}&
${\left\{\right\}}$
\\
WRAN$_{5}$&
{$\left\{ {\rm CH}_{2} \right\}$}&
{$\left\{ {\rm CH}_{5} \right\}$}&
{$\left\{\right\}$}&
{$\left\{\right\}$}&
${\left\{\right\}}$
\\
WRAN$_{6}$&
{$\left\{\right\}$}&
{$\left\{ {\rm CH}_{1} \cdots {\rm CH}_{10} \right\}$}&
{$\left\{\right\}$}&
{$\left\{\right\}$}&
${\left\{\right\}}$
\\
WRAN$_{7}$&
{$\left\{{\rm CH}_{7} \right\}$}&
{$\left\{\right\}$}&
{$\left\{\right\}$}&
{$\left\{\right\}$}&
${\left\{\right\}}$
\\
WRAN$_{8}$&
{$\left\{{\rm CH}_{1} \right\}$}&
{$\left\{{\rm CH}_{1} \right\}$}&
{$\left\{\right\}$}&
{$\left\{\right\}$}&
${\left\{\right\}}$
\\
WRAN$_{9}$&
{$\left\{{\rm CH}_{9} \right\}$}&
{$\left\{ \right\}$}&
{$\left\{{\rm CH}_{6} \right\}$}&
{$\left\{ {\rm CH}_{2} , {\rm CH}_{10} \right\}$}&
${\left\{\right\}}$
\\
WRAN$_{10}$&
{$\left\{{\rm CH}_{3} \right\}$}&
{$\left\{ \right\}$}&
{$\left\{{\rm CH}_{7} \right\}$}&
{$\left\{\right\}$}&
${\left\{{\rm CH}_{8} \right\}}$
\\
WRAN$_{11}$&
{$\left\{{\rm CH}_{7} \right\}$}&
{$\left\{ \right\}$}&
{$\left\{{\rm CH}_{5} \right\}$}&
{$\left\{{\rm CH}_{10} \right\}$}&
${\left\{{\rm CH}_{6} \right\}}$
\\
WRAN$_{12}$&
{$\left\{{\rm CH}_{5} \right\}$}&
{$\left\{{\rm CH}_{5} \right\}$}&
{$\left\{ \right\}$}&
{$\left\{\right\}$}&
${\left\{ \right\}}$ \\
\hline
\end{tabular}
\label{tab1}
\end{center}
\end{table*}

Collaborative sensing can be realized in a centralized or distributed fashion \cite{Ganesan2007I}. From a different perspective, existing cooperative sensing mechanisms are different in (i) the way sensors are selected and synchronized (sensor selection problem), (ii) what local sensing information is relayed (sensor relaying problem), and (iii) how these local results are fused (sensor fusion problem). Three approaches have been identified for sensor relaying: hard-decision combining (HDC), quantized soft combining, and soft-decision combining (SDC). While HDC and SDC have no distinctive edge over each other, quantized soft combining methods seek a trade-off between the two. In fact, while the local sensing decisions in HDC are binary (with 0/1 indicating the absence/presence of incumbent transmission), in SDC, any test statistic may be outputted. Next, we delve into these two classes in further detail.
\subsubsection{Hard-Decision Combining}
One of the shortcomings of the HDC schemes is their pure reliance on ``sensing outcomes" whereas ``sensing reliability" is as important. The significance of the latter lies beneath the reality that cognitive sensors operate in a wireless environment tangled with random phenomena that erratically diminish the credibility of the sensing outcomes \cite{Qihang2006}. The other shortcoming develops in collaborative schemes that neglect the existence of correlated shadowing, which lessens the diversity degree and dwindles the sensing performance \cite{Ghasemi2007}. Recognizing that both secondary and primary signal powers fall off with distance and that the frequency can be reused far from PUs, a distributed sensing mechanism was proposed in \cite{Bazerque2010} to derive a map for the power spectral density in frequency and space. This map can then be used by cognitive nodes to decide whether opportunistic transmission is safe or not, even though the primary transmission may be present. Static approaches such as AND, OR, VOTING, and linear quadratic combination \cite{Unnikrishnan2008} are among HDC methods.
\subsubsection{Soft-Decision Combining}
Among the SDC methods, fusion rules that are based on log-likelihood ratio test were proven to be optimal \cite{Varshney1986, Lunden2007, Zahabi2012}. Problems of this kind are mostly formulated as a classification problem \cite{Visotsky2005} and are computationally complex, especially in situations where the sensing decision is to be taken in real-time. Alternatively, there are other simpler methods to combine the individual sensing outcomes, including maximal ratio combining (MRC), equal gain combining (EGC) \cite{Hamza2014}, selection combining (SC) and switch and stay combining (SSC). It was also proven that HDC has a very close performance to SDC \cite{Mishra2006}, \cite{Hamza2014}.
\subsubsection{Sensing Fusion in IEEE 802.22}
Since the release of the WRAN standard, limited efforts have been dedicated to one of the most fundamental open issues pointed out in \cite{IEEE80222}, namely, sensing fusion. The WRAN SFMs proposed in \cite{Lim2009} operate based on the idea of exploiting the sensors' confidence metrics. \cite{Kim2010} presents a sensing clustering mechanism for in-band sensing, discusses the issue of ``how often to sense the channel?" in WRANs and proposes a scheduling mechanism for such. Despite its implementation complexity, the approach in \cite{Kim2010} alleviates some of the shortcomings associated with the ``OR" SFM. The core clustering idea in \cite{Min2011} is the same as \cite{Kim2010}, except that the former puts emphasis on security aspects of the mechanism by using attack-tolerant collaborative sensing. The closest study to the current paper is \cite{TadayonGC2015} which proposes a single-channel learning-based distributed sensing mechanism for WRANs operating in single-cell and normal operational mode. The present paper is a profound multi-channel extension of \cite{TadayonGC2015} to WRANs operating in coexistence mode.
\subsection{Channel Management}
The collaborative SFM proposed in  \cite{IEEE80222} requires CPEs (a.k.a. cognitive nodes or secondary users) to first sense their operating channels individually and, then, report the results to the base station (BS) intermittently. The sensing tasks should be carried out during synchronized quiet periods (${\rm QP}$s), whereby all CPEs shall go silent to reduce false-alarm events. The network-wide and periodic coordination of ${\rm QP}$s is one of the responsibilities of the BS. In this regard, different ${\rm QP}$ schedules may be required depending on the number of \textit{operating channels} being used by CPEs within a cell. Also, ${\rm QP}$s at different channels can be overlapping or non-overlapping. Once the sensing outcome of each CPE on the status of its operating channel is reported to its associated BS,\footnote{Hereinafter, sensor and CPE may be used interchangeably, anywhere the context is channel sensing.} it is the responsibility of the BS to make the decision upon the status of the sensed channels and to administer any respective action. To enhance the reporting reliability, \cite{IEEE80222} devises different notification and reporting mechanisms including contention-based, CDMA-based through urgent coexistent situation (UCS) window and bulk measurement messaging (BLM), as well as reporting through CPEs' allocated upstream bursts.

The rest of this section delineates the relevant WRAN functionality, whose knowledge is needed for understanding the sensing mechanism proposed in this paper. Before further progression, the following notation is introduced: subscripts $_{i}$, $_{j}$, $_{k}$ represent the CPE index, WRAN cell index (in a network  with $\mathfrak{M}$ cells) and the operating channel index (in a network with $\mathfrak{B}$ channels), respectively. Table \ref{tab0} consolidates the notations and acronyms used in this paper.
\subsubsection{Channel Switching}
Once the BS in ${\rm WRAN}_j$ cell concludes that one of its operating channels, say ${\rm CH}_{b}$,  is busy, ${\rm CH}_{b}$ and its two adjacent channels, ${\rm CH}_{b\pm 1}$, shall be vacated, removed from the {\it{operating channel list}} (OCL) and added to the \textit{protected channel list} (PCL), which is then followed by the high priority channel switching mechanism \cite[7.22]{IEEE80222}. Upon this change, the BS assigns a timer with long enough duration to ${\rm CPE}_{i,j}$ in ${\rm WRAN}_{j}$ to ascertain CPEs are prepared for the aforementioned switching operation, provided that such duration is no longer that the maximum allowable channel moving time. Subsequently, a channel from the \textit{backup channel list} (BCL) of ${\rm WRAN}_{j}$ is selected and the switching procedure gets initiated. This mechanism is called in-band sensing (IBS) \cite[10.3.3]{IEEE80222} because the sensing is carried out on the operating channels, as opposed to the out-of-band sensing (OBS) \cite[10.3.1]{IEEE80222} on the {\it protected}, {\it candidate} and {\it backup} channels \cite[10.2.3]{IEEE80222}.
\begin{figure}[t]
\includegraphics[scale=0.47]{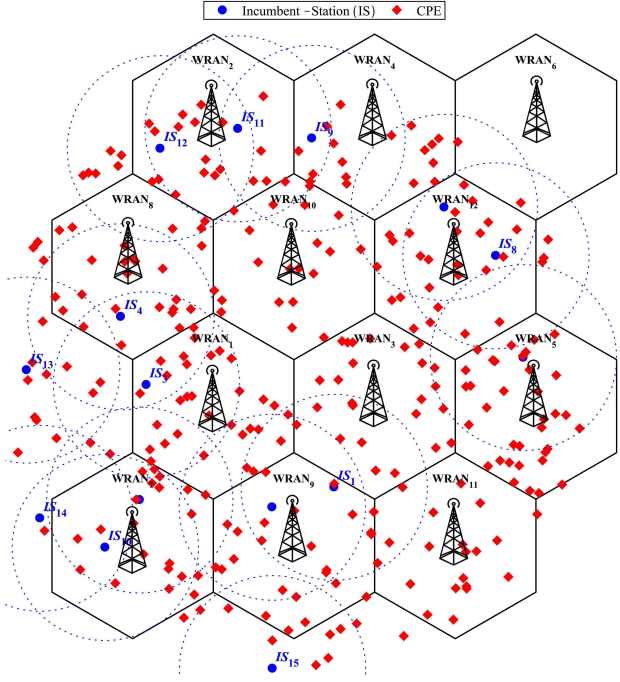}
\centering
\caption{An example of a WRAN with $\mathfrak{M}=12$ coexisting cells (${\rm WRAN}_j$) and CPEs uniformly distributed. A total of $15$ incumbent (primary) stations (IS) are shown with solid blue dots surrounded by dotted circles indicating their coverage.}
\label{fig:FigureI}
\end{figure}

\begin{figure*}[t]
\includegraphics[scale=0.5]{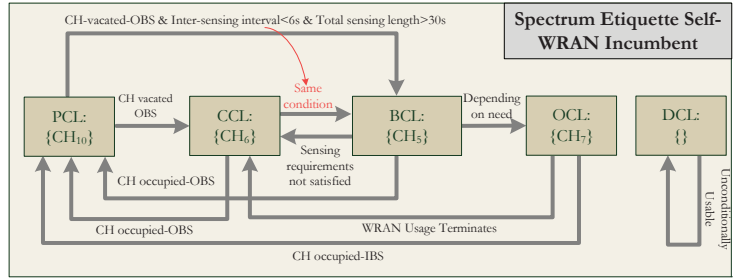}
\centering
\caption{IEEE 802.22 channel transition diagram for $\rm{WRAN}_{11}$ in the simulation scenario of Fig. \ref{fig:FigureI}.}
\label{fig:FigureII}
\end{figure*}

The importance of OBS is due to the vital need to always maintain a sufficient number of channels in the OCL, through a closed-loop mechanism (shown in Fig. \ref{fig:FigureII}) giving fair opportunities to all channels to return to the OCL as the incumbent transmission vanishes. Therefore, some CPEs are to be specifically selected and assigned to perform OBS on channels in BCL, PCL and the \textit{candidate channel list} (CCL). Since no payload is transmitted over these channels, they do not need to be sensed during network-wide ${\rm QP}$s, thus are less burdensome.

Notwithstanding the similarity between OBS and IBS reporting and decision-making procedures, these two mechanisms differ in the way the corresponding channel lists are updated. Specifically, a channel in PCL, say ${\rm CH}_{b}$, that is sensed idle during OBS is removed from this list and added to the CCL. Then, ${\rm CH}_b$ will be added to the BCL only if it has remained incumbent-free for no less than $30$ sec, with inter-sensing intervals not exceeding $6$ sec each. On the flip side, ${\rm CH}_b$ remains in PCL as long as it is sensed busy by OBS sensors. In the former case, and after added to the BCL, ${\rm CH}_b$ either remains still as long as the OBS flag is not raised (otherwise, it shall return to the PCL) or enters the OCL upon the channel switching event. The channel transitions between lists implemented according to the above rules are compactly represented in the form of a transition diagram (Fig. \ref{fig:FigureII}). Finally, \textit{disallowed channel list} (DCL) bears the indices of channels that are precluded from use by incumbents, thus, can be unconditionally used by WRAN. Table \ref{tab1} illustrates a state of the simulated WRAN shown in Fig. \ref{fig:FigureI}, where $12$ coexisting cells operate on $10$ channels.
%
%
\subsubsection{Channel List Updating}
\begin{figure}[b!]
\includegraphics[scale=0.95]{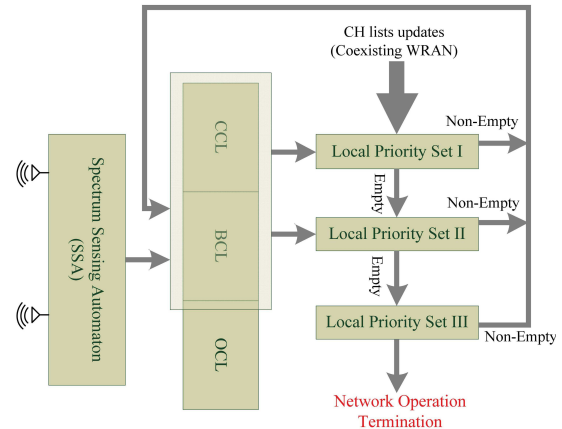}
\centering
\caption{Schematic illustration of the IEEE 802.22 backup CL and candidate CL updating mechanism in coexistence mode.}
\label{fig:FigureIII}
\end{figure}

When operated in coexistence mode (see Section II.C), neighbouring WRAN cells may share similar channel pools. In such situation, changes in BCL and CCL of ${\rm WRAN}_{j}$ may be triggered for two reasons: (i) a change in the status of the channels sensed in ${\rm WRAN}_{j}$, as described earlier, and (ii) a change in the channel lists of ${\rm WRAN}_{j}$'s neighboring cells  \cite[10.2.3]{IEEE80222}. This is due to the restrictions imposed by \cite{IEEE80222} on the choices of the operating and backup channels in neighboring cells. For instance, neighboring cells are not initially allowed to use similar operating and backup channels at the same time. Therefore, in case of channel swapping in the lists of ${\rm WRAN}_{j}$, neighboring cells may have to update their PCL and CCL lists, accordingly. For such updating, ${\rm WRAN}_{j}$ uses the following {\it local priority sets} (LPS),
\begin{equation}\label{Equation1}
\begin{split}
     {{\mathrm{LPS}_j}^\mathrm{I}}&=\left\{\mathrm{BCL}_{j} \cup \mathrm{CCL}_{j} \right\}  \setminus\bigcup \limits_{\mathit{l}\in \mathit(-j)}\left\{\mathrm{CCL}_{\mathit{l}}\cup \mathrm{BCL}_{\mathit{l}}\right\}
\\
      {\mathrm{LPS}_{j}}^\mathrm{II}&=\left\{\mathrm{BCL}_{j} \cup \mathrm{CCL}_{j} \right\} \setminus \bigcup \limits_{\mathit{l}\in \mathit(-j)}\left\{\mathrm{OCL}_{\mathit{l}}\right\}
\\
    {\mathrm{LPS}_{j}}^\mathrm{III}&=\bigcup \limits_{\mathit{l}\in \mathit(-j)}\left\{\mathrm{OCL}_{\mathit{l}}\right\}
\end{split}
\end{equation}
where the subscript $_{\mathit{l}}$  indicates the association with ${\rm WRAN}_{\mathit{l}}$. The operators $\setminus$ and $\cup$ represent the set difference (a.k.a. exclusion) and union, respectively, and $(-j)$ signifies that the union operation is to be carried out over all the neighboring cells of ${\rm WRAN}_{j}$ except itself. Since the updating decisions are to be made centrally by the BS, it utilizes sets in \eqref{Equation1} by their priorities, meaning ${\rm{LPS}_{j}}^{\rm{III}}$ is used for list updating only when ${\rm{LPS}_{j}}^{\rm{I}}$ and ${\rm{LPS}_{j}}^{\rm{II}}$ are both empty, whereas ${\rm{LPS}_{j}}^{\rm{II}}$ is used only if ${\rm{LPS}_{j}}^{\rm{I}}$ is empty (otherwise, ${\rm{LPS}_{j}}^{\rm{I}}$ is utilized). This CL updating mechanism is illustrated in Fig. \ref{fig:FigureIII}. In the end, the BS uses the superframe control header (SCH) to update the local OCLs of CPEs, and uses \textit{downstream channel descriptor} (DCD) to update the local PCLs and CCLs of CPEs \cite{IEEE80222}.
\subsection{Normal Mode vs. Coexistence Mode}
The IEEE 802.22 standard allows two modes of operation, namely, {\it normal} and {\it coexistence}. Unlike the normal mode of operation in which only a single WRAN exists to transmit in all the resource blocks (RBs) or multiple WRANs exist but with totally isolated frequency bands, in the coexistence mode \cite[7.20]{IEEE80222} one or more channels may be shared among neighboring WRAN cells. It is to be noted that the {\it coexistence beacon protocol} (CBP) is responsible to achieve synchronization among operating cells by transmitting control packets during {\it self-coexistence windows} (SCW). Readers are referred to \cite{Stevenson2009, Tadayon2013WRAN} to know more about WRAN frame structure.

The interference-free requirements in IEEE 802.22 WRANs not only necessitate that a channel discovered as busy and its two adjacent channels get immediately vacated to avoid inter-carrier interference, but also impose constraints on the choice of the operating channels in neighboring WRAN cells (in the coexistence mode). More narrowly, the CBP acts as a coordinator by prohibiting neighboring cells from using similar operating channel or by settling down a compromise.
\subsection{Quiet Period Management}
One of the prime challenges of IBS is its uncompromising requirement to silent all CPEs for a span of time, i.e. ${\rm QP}$. Despite their important role in accurate sensing, ${\rm QP}$s can severely degrade sensing efficiency and quality-of-service (QoS) since no data payload is transmitted on the operating channels in these interims. In the particular situation where \rm{inter-frame sensing} is obligated, which lasts for an entire superframe duration, the QoS damage can sometimes be even more acute \cite[7.21.1]{IEEE80222}. On the other hand, the OBS execution does not require network-wide ${\rm QP}$ scheduling as the target channels are not used for transmission, which ultimately burdens less sensing cost.

To deal with its inherent shortcoming and meet channel detection-time requirements \cite[7.21]{IEEE80222}, IBS in WRAN is carried out in two different scales: {\it intra-frame} IBS and {\it inter-frame} IBS. This two-stage ${\rm QP}$ management enables the network to adjust the sensing length and repetition rate, as per needs. In fact, the sensing procedure always starts with the more frequent, shorter (lasting for a frame length of $\cong 10{\rm ms}$) and less complex \rm{intra-frame IBS}, which may be followed by more precise, but longer \rm{inter-frame IBS} (lasting for upto a superframe length $\cong 160{\rm ms}$) utilizing more complicated algorithms, in case the detection requirements are not met by intra-frame IBS. When WRAN is operated in coexistence mode, both ${\rm QP}$ types shall be synchronized between neighboring cells using {\it superframe control header} (SCH) or channel request/response messages.
\section{The Proposed Multi-channel Sensing Fusion Mechanism}
As pointed out earlier, the standard left the choice of a ubiquitous SFM an open issue for most of the regulatory domains \cite[8.6.3.1]{IEEE80222}. Seemingly, this is due to the uncertainties that the relevant working groups bore about the performance optimality, efficiency and practicality of the existing distributed sensing solutions in the literature as well as their integrability within the standard's framework. Therefore, \cite{IEEE80222} proposed basic SFMs with deterministic performance, such as AND, OR and VOTING rules, as an interim solution until more fitting mechanisms appear. Moreover, \cite{IEEE80222} does not mandate the aforementioned SFMs for most of the regulatory domains, except for the United States, which certainly is an unduly cautious choice whose ramification is the waste of resources manifested by high false-alarm probability. All these became our motivations to work out an efficient, standard-compatible, and easily implementable, multi-channel learning-based distributed SFM: the proposed MC-LDS. Seeing what is proposed in this paper as a truly generic fusion scheme, the authors discern a broader range of applicability for MC-LDS beyond WRANs.

Let us focus on Fig. \ref{fig:FigureI}, which depicts a network layout consisting of $\mathfrak{M}=12$ coexisting cells within which CPEs are uniformly distributed. The incumbent base stations (IS) are represented with solid dots whose coverage zones are shown with overlapping dashed circles. A number of channels conceded to the IS are opportunistically accessed by WRAN cells. The proposed MC-LDS mechanism, reinforced from underneath by a reward-penalty rationality as well as two levels of differentiation, is detailed next.
\subsection{Decision Binarization}
As detailed in \cite{IEEE80222}, sensors should individually sense the channels within synchronized ${\rm QP}$s. As explained earlier, in combining the sensing outcomes, HDC methods are less complex but less optimal compared to SDC methods. To have the best out of both worlds, any future SFM proposal for amendment to the standard is better to be a combination of HDC and SDC.  This is the basic idea behind MC-LDS. Herein, energy detection is adopted for it is analytical tractability and its prevalence. Sensing by energy detection requires the least processing, which is basically done by  taking $\mathcal{M}$ discrete samples from the received signal $y$ and forming the power-sum $\mathcal{S}=\sum_{j=1}^{\mathcal{M}}{|y[j]|^2}$. The major drawback of sensing by energy detection is its vulnerability to noise uncertainties, which can be tackled through cooperation.

\begin{figure}[b]
\centering
\subfloat[Misdetection and false-alarm trade-off.]{\includegraphics[scale=0.45]{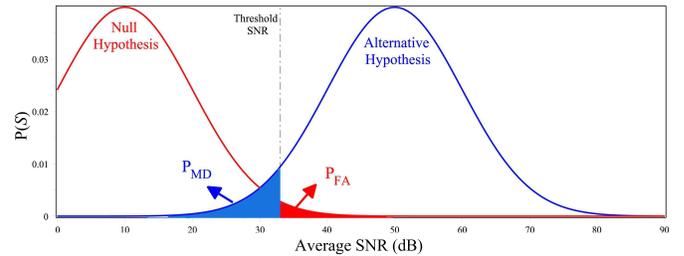} \label{fig:FigureIVa}}\\
\subfloat[Logisitc classifier.]{\includegraphics[scale=0.45]{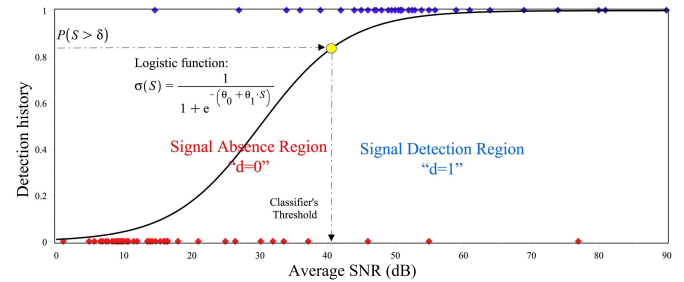} \label{fig:FigureIVb}}
\caption{Sensor's decision binarization problem to discover the status ($d$) of its operating channel.}
\label{fig:FigureIV}
\end{figure}

Irrespective of the type, the performance of any spectrum sensing mechanism is measured by four rates: (1) true-positive, (2) true-negative, (3) false-positive (a.k.a. false-alarm rate, $P_{\rm FA}$), and (4) false-negative (a.k.a. misdetection rate, $P_{\rm MD}$). It is to be noted that the consolidation of (1) and (2) into a single quantity is also recognized as successful discovering probability, $P_{\rm SD}$. The false-alarm probability, $P_{\rm FA}$, quantifies the sensor's misperception acuity in detecting a primary transmission that does not actually exist. On the other hand, the misdetection rate, $P_{\rm MD}$, denotes the sensor's inability to detect a primary transmission that does actually exist. Hereafter, the superscript $^{(t)}$ over symbols denotes the ${\rm QP}$ index, which may be signified by ${\rm QP}^{(t)}$, as well. Adhering to this notation system, and recalling that indices $i$, $j$ and $k$ denote the sensor, WRAN cell and operating channel, respectively, the aforementioned quantities are mathematically expressed as:
\begin{equation}\label{Equation2} \nonumber
\hspace{-0.3 cm}
\begin{array}{l l}
\begin{array}{l}
P_{{\rm MD},i,j}^{(t),k}=\Pr\left(\mathcal{E}_1 \right)\\
P_{{\rm FA},i,j}^{(t),k}=\Pr\left(\mathcal{E}_0 \right)\\
P_{{\rm D},i,j}^{(t),k}=\Pr\left( \neg \mathcal{E}_1\lor \neg \mathcal{E}_0\right)
\end{array}, &
\begin{array}{l}
\hspace{-0.4 cm}
\mathcal{E}_1=\{\mathcal{S}_{i,j,k}^{(t)}<\lambda_{j,k} \, \vert \mathcal{H}_{1,k}^{(t)}\} \\
\hspace{-0.4 cm}
\mathcal{E}_0=\{\mathcal{S}_{i,j,k}^{(t)}>\lambda_{j,k} \, \vert \mathcal{H}_{0,k}^{(t)}\}
\end{array}
\end{array}
\end{equation}
where $\mathcal{H}_{1,k}^{(t)}$ ($\mathcal{H}_{0,k}^{(t)}$) are the alternative (null) hypotheses that the $k^{\rm th}$ channel is taken (not taken) by the IS, and where $\mathcal{S}_{i,j,k}^{(t)}$ is the collected power at the sensor's receiver and $\lambda_{j,k}$ its corresponding power threshold. As shown in Fig. \ref{fig:FigureIVa}, $P_{{\rm MD},i,j}^{(t),k}$ ($P_{{\rm FA},i,j}^{(t),k}$) can be geometrically interpreted as the area confined under two different received power $\mathcal{S}_{i,j,k}^{(t)}$s' probability distribution functions (PDFs) and delimited from left (right) to the ${\rm SNR}$ threshold  $\lambda_{j,k}/\mathcal{S}_N$, where $\mathcal{S}_N$ is the noise power.

The MC-LDS distributed SFM works with local binary sensing decisions (denoted $d^{(t)}_{i,j,k}$). Each $d^{(t)}_{i,j,k}$ is the solution of a binary classification problem that the sensor has to repeatedly solve after ${\rm QP}^{(t)}$. Though this puts MC-LDS somewhere inside the HDC class, here local decisions $d_{i,j,k}^{(t)}$ are not made statically using an unrefined strategy such as $\Pr({\mathcal{S}}_{i,j,k}^{(t)}\geq \lambda_{j,k}) \to d_{i,j,k}^{(t)}=1$ (or $\Pr({\mathcal{S}}_{i,j,k}^{(t)}< \lambda_{j,k}) \to d_{i,j,k}^{(t)}=0$), but rather adaptively.

To solve the local classification problem, the logistic regression method is adopted. This is because logistic regression's outcome is the best fitting predictor among other classification supervised learning methods,\footnote{Example of other learning approaches are Bayesian networks, support vector machines, neural networks, \textit{Probit} regression and Naive-Bayes learning.} that not only forecasts the operating channel status for future moments, but its value is a meaningful continuum which can be interpreted as the probability that the operating channel is busy. Also, it is a well-accepted fact that logistic estimator is \textit{consistent} (chooses the right estimator in the limit when the number of training examples is large) and \textit{efficient} (no other estimator converges faster, in the mean squared sense). These traits make logistic regression the appropriate classification tool in problems where there is uncontrollable stochasticity in the system. Dubbed logistic (\textit{Sigmoid}) function, $0\leq\sigma\left(\mathcal{S}\right)<1$ is defined as the probability that the measured channel is busy and is characterized by,
\begin{equation}
\label{Equation3}
\sigma\left(\mathcal{S}=\mathcal{S}^{(t)}_{i,j,k}\right)=\displaystyle\frac{1}{1+\displaystyle e^{-(\theta_{0}+\theta_{1}\cdot\mathcal{S})}}\bigg|_{\mathcal{S}=\mathcal{S}^{(t)}_{i,j,k}}.
\end{equation}
The $i^{\rm th}$ sensor in  ${\rm WRAN}_{j}$ (hereinafter, represented as ${\rm Sen}_{i,j}$) uses $\sigma\left(\mathcal{S}\right)$ in the following maximum likelihood estimation (MLE) problem,  to find its corresponding regression parameters, $\Theta_{i}=[\theta_{0,i}$,$\theta_{1,i}]$:
%
\begin{equation}\label{Equation4}
\begin{split}
 \underset{{\Theta_{i}}}{\text{maximize}} \log \left( \prod\limits_{t\in \mathcal{N}_i}{\mathcal{L}\left(\sigma\left(\mathcal{S}_{i,j,k}^{(t)}\right) \big| {\Theta_{i}}\right)}\right), \\ 
 0\leq i\leq m_{j,k}; j\in \mathbb{M}=\{1\cdots \mathfrak{M}\}; k \in \mathbb{B}=\{1\cdots \mathfrak{B}\},
\end{split}
\end{equation}
where $m_{j,k}$ is the total number of CPEs in ${\rm WRAN}_{j}$ operating on ${\rm CH}_{k}$, $\{(\mathcal{S}_{i,j,k}^{(t)}, d_{i,j,k}^{(t)})\}^{t\in \mathcal{N}_i}$ is the set of training samples to be provided apriori to learn the classifier's parameters ($\Theta_{i}$), $\mathcal{S}_{i,j,k}^{(t)}$ is the  $t^{\rm th}$ measured power in $\mathcal{N}_i$, and $\mathcal{L}(\cdot)$ is the likelihood function to be evaluated at all training samples $\mathcal{N}_i$. The sensing outcomes $d_{i,j,k}^{(t)}\in\{-1,1\}$ being binary, the likelihood function $\mathcal{L}(\cdot)$ is in the form of a Bernoulli distribution:
%
\begin{equation}\label{Equation5}
\small \mathcal{L}\left(\sigma\left(\mathcal{S}_{i,j,k}^{(t)}\right)\big|{\Theta_{i}}\right)=\left(\sigma\left(\mathcal{S}_{i,j,k}^{(t)}\right)\right)^{d_{i,j,k}^{(t)}}\left(1-\sigma\left(\mathcal{S}_{i,j,k}^{(t)}\right)\right)^{1-d_{i,j,k}^{(t)}}.
\end{equation}

Quite obvious from \eqref{Equation5}, $d=1 (\text{or}~0)\to \mathcal{L}(\mathcal{S})=\sigma(\mathcal{S}) (\text{or } 1-\sigma(\mathcal{S}))$. The MLE in (\ref{Equation4}) is convex in $\Theta_i$. Hence, a unique optimal  $\hat{\Theta}_{i}$ exists, which can be found efficiently in polynomial time. Once obtained, this optimal $\hat{\Theta}_{i}$ is plugged into  (\ref{Equation3}) giving the corresponding channel busyness probability, i.e., $\sigma\left(\mathcal{S}_{i,j,k}^{(t)}=x\right)=\Pr\left(d_{i,j,k}^{(t)}=1\big|\mathcal{S}_{i,j,k}^{(t)}=x\right)$. Such probability may either be used directly  (SDC) or be binarized (HDC) in the decision combination step. We adopt the latter approach in subsection III.B. To that end, and since the predictor $\sigma\left(\mathcal{S}_{i,j,k}^{(t)}=x\right)$ is a continuous function in $[0,1]$,  $\forall x\in [-\infty,\infty]$, as depicted in Fig. \ref{fig:FigureIVb}, the following discriminant rule is applied,
\begin{equation}\label{Equation6}
d^{(t)}_{i,j,k}=\mathbbm{1}_{\mathcal{A}_{i}}\left(\mathcal{S}^{(t)}_{i,j,k}=x\right)=\left\{
     \begin{array}{lr}
       1 & \text{if } x \in \mathcal{A}_{i}\\
       0 & \text{if } x \notin \mathcal{A}_{i}
     \end{array}
   \right. ,
\end{equation}
where $\mathbbm{1}_{\mathcal{A}_{i}}(\cdot)$ is the Heaviside step function.

To duly characterize \eqref{Equation6}, sub-interval $\mathcal{A}_{i}$ is to be identified. In the case of logistic regression, this task is not so difficult as the predictor function $\sigma\left(\mathcal{S}\right)$ is monotonically increasing and, hence, $\mathcal{A}_{i}$ is a convex lower-bounded interval. Therefore, as shown in Fig. \ref{fig:FigureIVb} (yellow circle on the curve), if one could manage to know the probability $\Pr\left(\mathcal{S}^{(t)}_{i,j,k}>\lambda_{j,k}\right)$, then the corresponding power obtainable from inverting the predictor function \eqref{Equation3} at this probability would be the aforementioned $\mathcal{A}_{i}$'s lower bound. Mathematically stated, $\mathcal{A}_{i}=\left[\sigma^{-1}\left(\Pr\left(\mathcal{S}^{(t)}_{i,j,k}>\lambda_{j,k}\right)\right),\infty\right]$. Fortunately,  $\Pr\left(\mathcal{S}^{(t)}_{i,j,k}>\lambda_{j,k}\right)$ for sensor ${\rm Sen}_{i,j}$ in ${\rm CH}_{k}$ at ${\rm QP}^{(t)}$ can be obtained from the following marginalization,
%
%
\begin{equation}\label{Equation7}
\begin{split}
 \Pr \left(\mathcal{S}^{(t)}_{i,j,k}>\lambda_{j,k}\right) =\sum\limits_{l=0}^{1}{\Pr\left(\mathcal{S}^{(t)}_{i,j,k}>\lambda_{j,k}\big|H^{(t)}_{l,k}\right)\Pr\left(H^{(t)}_{l,k}\right)}&\\
=1-P^{(t),k}_{{\rm MD},i,j}+\Pr\left(H^{(t)}_{0,k}\right)\left(P^{(t),k}_{{\rm FA},i,j}+P^{(t),k}_{{\rm MD},i,j}-1\right),&
\end{split}
\end{equation}
which depends on the misdetection and false-alarm rates. There are two methods to know these rates: the first is statistical, more precise and requires some test samples $\left\{\left(\mathcal{S}^{(t)}_{i,j,k},d^{(t)}_{i,j,k}\right)\right\}^{t \in \mathcal{G}_i}$  (where the test sample set $\mathcal{G}_i$ shall be distinct from the training sample set $\mathcal{N}_i$) in order to compute antecedent empirical rates by counting false-positive and false-negative instances; otherwise, less precise approximate analytical expressions, such as the following one derived in  \cite{Tadayon2015MIMC}, may be used
\begin{equation}\label{Equation8}
\begin{array}{l}
P^{(t), k}_{{\rm MD},i,j}\cong 1-\mathbb{Q}\left(\dfrac{{\rm SNR}^{{\rm min}}_{j,k}-\dfrac{1+{\rm SNR}^{(t)}_{i,j,k}}{\beta^{(t)^{\rm Sen}}_{i,j,k}}\mathcal{M}_i}{\sqrt{\mathcal{M}_i\left(\left(\dfrac{{\rm SNR}^{(t)}_{i,j,k}}{\beta^{(t)^{\rm Sen}}_{i,j,k}}+2\right)^2-2\right)}}\right),\\\\
P^{(t), k}_{{\rm FA},i,j}\cong \mathbb{Q}\left(\dfrac{{\rm SNR}^{{\rm min}}_{j,k}-\mathcal{M}_i}{\sqrt{2\mathcal{M}_i}}  \right),
\end{array}
\end{equation}
where ${\rm SNR}^{{\rm min}}_{j,k}$ and ${\rm SNR}^{(t)}_{i,j,k}$ denote ${\rm Sen}_{i,j}$'s minimum and instantaneous ${\rm SNR}$, respectively, and $\mathbb{Q}(\cdot)$ is the Normal Q-function. Moreover, $\beta^{(t)^{\rm Sen}}_{i,j,k}$ and $\mathcal{M}_i$ represent the sensing channel's loss/fading coefficient\footnote{This is to be contrasted from the reporting channel coefficient $\beta^{(t)^{\rm Rep}}_{i,j,k}$, as will be encountered later on.} and the number of samples taken from the received signal, respectively. As pointed out before, ${\rm SNR}^{{\rm min}}_{j,k}=\lambda_{j,k}/\mathcal{S}_N$, where $\mathcal{S}_N$ is the noise power and $\lambda_{j,k}$ is the power level threshold.

Having binarized the local sensing outcomes $d^{(t)}_{i,j,k}$, the next step is to combine them to find an optimum global decision. Before tackling the decision combining problem, it should be noted that the MLE in (\ref{Equation4}) does not have to be solved at every ${\rm QP}$. Instead, it may be worked out once enough time is elapsed since the last estimation, or upon the availability of new strong evidences substantiating that the predictor's error is unacceptable.
\subsection{Decision Combination}
The aptness of the MC-LDS SFM hinges on two internalized levels of differentiation: temporal and spatial. Before describing the temporal differentiation, we need to introduce the reward-penalty logic used by MC-LDS. The logic here is to use the following rules to affix different scores (positive or negative) to sensors depending on situations that may come up: at ${\rm QP}^{(t)}$, given the set $\{d_{i,j,k}^{(t)}\}_{i=0}^{m_{j,k}}$ of the most recent sensing measurements from $m_{j,k}+1$ sensors participating in sensing ${\rm CH}_{k}$ in ${\rm WRAN}_{j}$,\footnote{Index $i=0$ is used to denote the BS of ${\rm WRAN}_{j}$ sensing ${\rm CH}_{k}$ since, according to \cite{IEEE80222}, all WRAN BSs are self-appointed to carry out separate sensing task alongside the CPEs. Thus, $m_{j,k}+1$ sensors sense the channel at anytime.} the latest central sensing decision $D^{(t-1)}_{j,k}$, and given the channel status estimate $R^{(t)}_{j,k}$ obtainable from the database service, the BS assigns the following rewards-penalty scores ${L}_{i,j,k}^{(t)}$ to $m_{j,k}+1$ participating sensors, $ \forall j\in\mathbb{M}, k\in\mathbb{B}$, and provided that $0<\gamma_{k}<\zeta_{k}$:
\begin{equation}\label{Equation9}
\begin{split}
&{{L}_{i,j,k}^{{(t)}^{\rm rew}}}=\left\{ {\begin{array}{l}
\gamma_{k}  \quad \text{if } d_{i,j,k}^{( t) }=R^{\left( t \right)}_{j,k}~\wedge~  d_{i,j,k}^{\left( t \right)}=D^{\left( t-1 \right)}_{j,k} \\
\zeta_{k}  \quad \text{if } d_{i,j,k}^{\left( t \right)}=R^{\left( t \right)}_{j,k}~\wedge~d_{i,j,k}^{\left( t \right)} \ne D^{\left( t-1 \right)}_{j,k}, \\
\end{array}} \right.\\ &
{{L}_{i,j,k}^{{(t)}^{\rm pen}}}=\left\{ {\begin{array}{l}
-\zeta_{k}  \quad \text{if }  d_{i,j,k}^{\left( t \right)}\ne R^{\left( t \right)}_{j,k}~\wedge~d_{i,j,k}^{\left( t \right)}=D^{\left( t-1 \right)}_{j,k} \\
-\gamma_{k}  \quad \text{if }  d_{i,j,k}^{\left( t \right)}
\ne R^{\left( t \right)}_{j,k}~\wedge~d_{i,j,k}^{\left( t \right)} \ne D^{\left( t-1 \right)}_{j,k}
\end{array}} \right.
\end{split}
\end{equation}

\begin{figure}[t]
\includegraphics[scale=0.35]{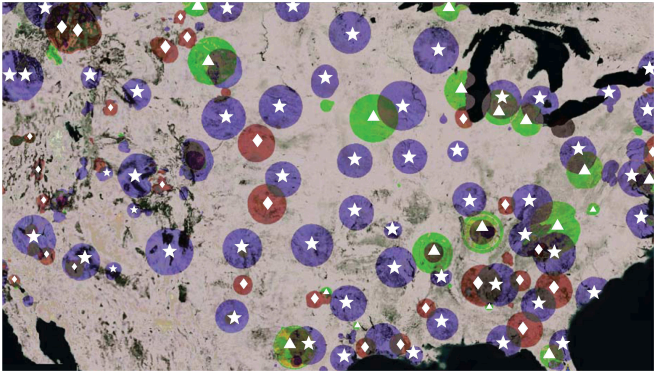}
\centering
\caption{Spatial footprint of incumbents occupying TV channels $k=5$ (green with centered triangles), $k=6$ (red with centered diamonds), and $k=7$ (violet with centered stars)  in USA, collected from \textit{White Space Plus} database (source: www.spectrumbridge.com).}
\label{fig:FigureV}
\end{figure}

Participation of database reading $R^{(t)}_{j,k}$ in \eqref{Equation9} is not arbitrary. This information shall be utilized toward enhancing the sensing precision and reliability as repeatedly emphasized and mandated in the standard \cite[10.2.2]{IEEE80222}. Accordingly, WRAN BSs should  retrieve the available channel list as well as the maximum allowed EIRP (effective isotropic radiated power) by sending queries to an incumbent database. Operated by third parties (c.f. FCC R\&O 08-260 \cite{FCC08260}), the service provided by such database is only a crude estimate of whether the queried channel is occupied at a given location-time or not. Concretely, factors such as passiveness of primary receivers (causing hidden terminal effect), non-real time database availability, and the unpredictable and varying nature of communication channel, make mere reliance on database service futile. Fig. \ref{fig:FigureV} is a spatial snapshot from a database wherein disks with different colors denote the keepout footprint of three channels $k=5,6,7$, where $R_{k}^{(t_0)}=1$  and WRAN subscript $j$ is omitted. Readers are referred to [1, subclause 10.2.2] for a more detailed treatment of the subject.

Since transmissions can take place on more than one operating channel, a cell can possess multiple values for $R^{(t)}_{j,k}$ and $D^{(t-1)}_{j,k}$ at any moment. Back to (\ref{Equation9}), the designation of the reward-penalty scores gives the highest reward ($\zeta_k$) to ${\rm Sen}_{i,j}$ whose decision accords with the database reading ($d^{(t)}_{i,j,k}=R^{(t)}_{j,k})$ but differs with the latest central decision ($D^{(t-1)}_{j,k}\ne R^{(t)}_{j,k}$). On the other hand, the highest penalty ($-\zeta_k$) is imposed to a sensor when the database reading neither agrees with the the latest central nor with that sensor's local decision. Table \ref{tab2} is a representation of (\ref{Equation9}) which, equivalently, reduces to the following mathematical expression:
\begin{equation}\label{Equation10}
\begin{split}
\hspace{-5pt}
L_{i,j,k}^{(t)}=& \gamma_k\left(d_{i,j,k}^{\left( t \right)}\odot R^{\left( t \right)}_{j,k}-d_{i,j,k}^{\left( t \right)} \oplus R^{\left(t
\right)}_{j,k}\right)\\-&
(\zeta_k-\gamma_k)\left[{\left(\neg d_{i,j,k}^{\left( t \right)}\right)}-d_{i,j,k}^{\left( t \right)} \right]\times \\&
\left[\left(\neg D^{\left( t-1 \right)}_{j,k}\right)R^{\left( t \right)}_{j,k}-D^{\left( t-1\right)}_{j,k}\left(\neg R^{\left( t \right)}_{j,k}\right) \right],\\
& \hspace{2.5 cm} j\in \mathbb{M}, k \in \mathbb{B}, 0\leq i \leq m_{j,k}
\end{split}
\end{equation}
where $\oplus$, $\odot$ and $\neg$ represent the XOR, XNOR and negation operators, respectively. Note that the total number of sensors in this cell is $m_{j}=\sum \limits_{k \in \mathbb{B}}{m_{j,k}}$ and that $m_{j,k}\ne 0$, $\forall k \in \mathrm{OCL}_{j}$ (operating channel list).

\begin{table}[h]
\begin{center}
\begin{small}
\caption{Reward-penalty scores in the MC-LDS mechanism.}
\begin{tabular}{p{15pt}|p{25pt}|p{15pt}|p{20pt}||p{15pt}|p{25pt}|p{15pt}|p{20pt}}
$d_{i,j,k}^{(t)}$&
$D^{(t-1)}_{j,k}$&
$R^{(t)}_{j,k}$&
${{L}_{i,j,k}^{{(t)}^{\rm rew}}}$&
$d_{i,j,k}^{(t)}$&
$D^{(t-1)}_{j,k}$&
$R^{(t)}_{j,k}$&
${{L}_{i,j,k}^{{(t)}^{\rm pen}}}$ \\[0.1cm]
\hline
0&0&0&$\gamma_k$&1&0&0&$-\gamma_k$ \\
1&0&1&$\zeta_k$&0&0&1&$-\zeta_k$\\
0&1&0&$\zeta_k$&1&1&0&$-\zeta_k$ \\
1&1&1&$\gamma_k$& 0&1&1&$-\gamma_k$
\end{tabular}
\label{tab2}
\end{small}
\end{center}
\end{table}

Continuing on, the first level of differentiation in MC-LDS is \textit{temporal}. The key idea is to allow both most recent and historical sensing results of ${\rm Sen}_{i,j}$ to participate in the decision-making process. However, to account for the fact that recent measurements are more dependable compared to older ones,\footnote{The existing coherence in the intensity of channel fading as well as the recurring transmission pattern of PUs are two underpinning reasons to sustain temporal differentiation.} the temporal discount tensor $\left[\alpha_{j,k}\right]$ is introduced, which may be chosen adaptively or statically. Therefore, by merging the reward-penalty scores, the confidence metric, $w_{i,j,k}^{(n)}$,\footnote{Mandated in \cite[10.4.1.2]{IEEE80222} as one of the spectrum sensing function outputs.} is derived for ${\rm Sen}_{i,j}$ at the present sensing interval, meaning ${\rm QP}^{(t=n)}$, according to
\begin{equation}\label{Equation11}
\begin{split}
w_{i,j,k}^{\left( n \right)}=\sum\limits_{t=n-N_{j,k}}^{n-1} {\alpha_{j,k}^{n-t}~L_{i,j,k}^{(t)}}, \hspace{1cm}  j\in \mathbb{M}, k \in \mathbb{B},
\end{split}
\end{equation}
where the superscript $^{n-t}$ is the mathematical exponentiation, as opposed to the indexing superscript $^{(t)}$. The historic count tensor $\left[N_{j,k}\right]$ represents the effective temporal length. As discussed in the next section, we implement static selection as well as dynamic adjustment when $\left[N_{j,k}\right]$ and $\left[\alpha_{j,k}\right]$ adaptively vary.  An implication of \eqref{Equation11} is that ${\rm Sen}_{i,j}$ that unceasingly inputs wrong information to its BS possesses a smaller (more negative) confidence metric than one with a more truthful record.

In the next step,  a discrete-to-continuous channel indicator $X_{i,j,k}^{(n)}$ is derived for ${\rm Sen}_{i,j}$ at the present sensing interval using its confidence metric $w_{i,j,k}^{(n)}$ and its sensing result $d_{i,j,k}^{(n)}$, according to
\begin{equation}\label{Equation12}
\begin{split}
X_{i,j,k}^{\left(n \right)}=\left\{ {\begin{array}{ll}
 w_{i,j,k}^{(n)} \quad {\rm if} \quad d_{i,j,k}^{(n)}=1 \\\\
 -w_{i,j,k}^{(n)} \quad {\rm if} \quad d_{i,j,k}^{(n)}=0 \\
 \end{array}} \right.,  \begin{array}{c}
   0\le i \le m_{j,k},\\
   j \in\mathbb{M}, \\
   k \in \mathbb{B}.
 \end{array}
\end{split}
\end{equation}

\begin{figure*}[tp]
\includegraphics[width=4.5in,height=2.6in]{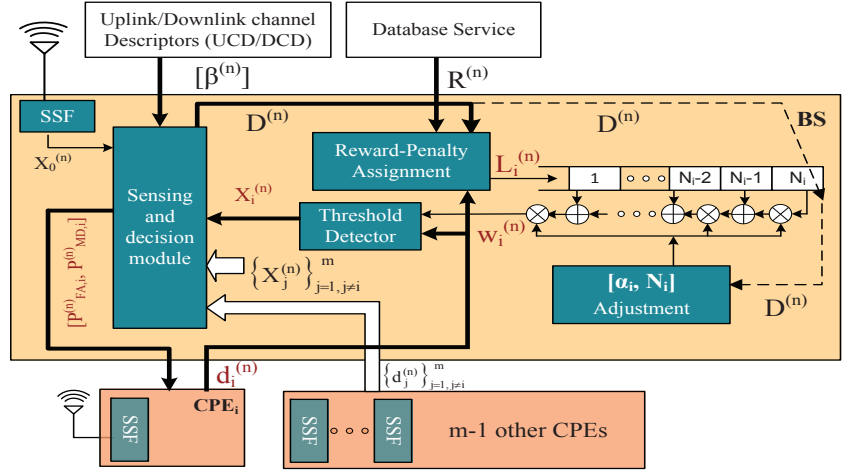}
\centering
\caption{The MC-LDS block diagram (cell/channel indices of symbols are removed for better clarity).}
\label{fig:FigureVI}
\end{figure*}

So far, the mechanism has dealt with the sensing outputs from sensors in solidarity. To make an intelligent decision and improve the reliability, these local sensing outcomes are to be combined. In the context of WRANs, after the vectors $\left[ d_{i,j,k}^{(n)},w_{i,j,k}^{(n)},X_{i,j,k}^{(n)} \right]$ are reported to the BS, it is the responsibility of this BS to make a central decision about the status of the operating channel ${\rm CH}_{k}$.

When combining outcomes, there is another source of error, which may degrade the sensing performance. More narrowly, while the sensing reliability of ${\rm Sen}_{i,j}$ is an important factor, the quality of the CPE-to-BS link, known as reporting channel, should also be taken into account. Thereby, in combining the local sensing outcomes, the second level of differentiation, known as \textit{spatial}, is sustained by dedicating larger trust on sensors whose reporting channel gains $\beta^{(n)^{\rm Rep}}_{i,j,k}$ are larger,
\begin{equation}\label{Equation13}
\begin{split}
D_{j,k}^{(n)}=\left\{ {\hspace{-0.2 cm} \begin{array}{l}
 1,  X_{0,j,k}^{(n)}+\displaystyle\sum\limits_{i=1}^{m_{j,k}} {\beta^{(n)^{\rm Rep}}_{i,j,k} X_{i,j,k}^{(n)}} >0 \\
 0, X_{0,j,k}^{(n)}+\displaystyle\sum\limits_{i=1}^{m_{j,k}} {\beta^{(n)^{\rm Rep}}_{i,j,k} X_{i,j,k}^{(n)}} \le 0 \\
 \end{array}} \right.,
 \begin{array}{l}
   j \in\mathbb{M} \\
   k \in \mathbb{B}
 \end{array}
\end{split}
\end{equation}
where $m_{j,k}$ is the number of sensors sensing ${\rm CH}_{k}$ in ${\rm WRAN}_{j}$, with the property  $m_{j,k}\ne 0~\forall k \in \mathrm{OCL}_{j}\neq \varnothing $, and $\beta^{(n)^{\rm Rep}}_{i,j,k}$ is the fading coefficient of the reporting channel for ${\rm Sen}_{i,j}$. Once again, the confidence metric $ X_{0,j,k}^{(n)}$ is to account for the involvement of the BS in the sensing process as an independent sensing entity.

According to (\ref{Equation13}), a more accurate sensor (larger $X_{i,j,k}^{(n)}$) able to deliver its measurement more reliably to the BS (larger $\beta^{(n)^{\rm Rep}}_{i,j,k}$) contributes more constructively in the final fusion process, compared to a less reliable sensor (smaller  $X_{i,j,k}^{(n)}$) or one with an erroneous reporting channel (smaller $\beta^{(n)^{\rm Rep}}_{i,j,k}$).

There is another important implication about the way (\ref{Equation12}) is defined in conjunction with (\ref{Equation13}). More precisely, a positive $X_{i,j,k}^{(n)}$ , which strengthens the possibility of declaring a busy channel in \eqref{Equation13}, is attained when one of these two cases occurs in (\ref{Equation12}):
(i) $\left[w_{i,j,k}^{(n)}\geq 0,d_{i,j,k}^{(n)}=1\right]$ and (ii) $\left[w_{i,j,k}^{(n)}<0,d_{i,j,k}^{(n)}=0\right]$. While (i) implies that a trustworthy sensor, which perceives the channel as busy in ${\rm QP}^{(t=n)}$ (i.e. current sensing interval), chips in directly in declaring a busy channel in \eqref{Equation13}, the direct implication from (ii) is that an unreliable sensor, which incorrectly discerns the channel status as idle, was forced to make a reverse influence. In other words, with MC-LDS, both unreliable and reliable sensors are identified based on their historical records and are constructively leveraged towards a more accurate final decision. Similar deductions are valid for negative $X_{i,j,k}^{(n)}$ resulting from two other cases: (iii) $\left[w_{i,j,k}^{(n)}<0,d_{i,j,k}^{(n)}=1\right]$ and (iv) $\left[w_{i,j,k}^{(n)}\geq 0,d_{i,j,k}^{(n)}=0\right]$.

The other important attribute of MC-LDS is the relativity of sensors' reliability metric (i.e. confidence metrics) that emerges from the fact that $X_{i,j,k}^{(n)}$ takes values from a continuum. The larger $\zeta_k$ and $\gamma_k$ are chosen in \eqref{Equation9}, the wider this continuum. This creates a differentiating property inherited form the designated rewarding-penalty scores in Table \ref{tab2}.

The block diagram in Fig. \ref{fig:FigureVI} summarizes the proposed SFM: MC-LDS.
\section{Simulation and Validation Results}
\subsection{Simulator Platform}
To verify the performance of MC-LDS, a discrete-even simulator was built in C++ and Maple programming languages. In the simulator, all the relevant management and controlling functions of WRAN were implemented. This chiefly includes the ${\rm QP}$ and sensing management procedures (including inter-frame and intra-frame sensing, synchronization and reporting), channel management functions (including channel list updating, channel switching scheduling, etc.), and incumbent protection module. Coexistence among WRAN cells was taken into accout through the implementation of the CBP protocol.

The topology in each cell is realized in ${\rm 2D}$ space, meaning that CPEs are assumed to be at the same height, which is close to reality for rural areas where WRAN antennas are normally mounted few meters above the ground level at almost the same elevation. All the communication settings, including transmit power, WRAN cell's diameter, ${\rm SNR}$ threshold, incumbent activity pattern, channel width, interference temperature, maximum allowed interference time, etc. were chosen based on recommendations and mandates of \cite{IEEE80222} (in particular \textit{Annex A}) and, if not alluded, typical values were chosen. The channel degrading factors, noise and fading, were implemented for both the sensing and the reporting channels. In addition, correlated channel fading \cite{Gudmundson1991} was taken into account, by restricting CPEs that happen to be in $\lambda /4$ vicinity to experience similar channel gains from the BS.  User mobility was not considered in the simulator since WRANs are aimed for static household rather than mobile users. This is apart from the fact that larger antenna size in UHF/VHF frequencies poses serious practical considerations for mobile applications. For the fact that CPEs are immobile, population density is low, and reflectors, refractors and scatterers are proportionately motionless, slow fading is a reasonable model for temporal variations of the wireless channel in rural areas. The latter was implemented by affixing similar channel gains over several consecutive time frames. Further, the incumbent activity pattern is assumed to be independent across multiple  available channels, where the activity ratio (i.e. ${\rm IAR}$=ON/OFF) on each channel is a mix of bursty and non-bursty patterns with the possibility to adjust the ${\rm IAR}$.

At the simulation start-up, a number of channels are randomly distributed among ISs, where two or more different ISs may broadcast the same channel to their respective users. In the latter case, the footage of two or more ISs may overlap to have the whole geographical area under coverage of the incumbent service. This may lead to a CPE accidentally located within the transmission range of two or more active ISs to collect a  primary signal with a large superimposed power from all  sources, an incidence that can influence the false-alarm and misdetection rates, significantly, but neglected in almost all models. This effect was also properly realized in the simulator.
\subsection{Performance Measures}
To evaluate the performance of MC-LDS, several performance measures are introduced in this paper.
\subsubsection{False Alarm Rate}
Theoretically defined as in (\ref{Equation2}), the false-alarm rate can be experimentally estimated, for an individual sensor as well as for the entire cell, using the following formulae,
\begin{equation}\label{Equation14}
\begin{split}
\begin{array}{c}
P_{{\rm FA},i,j}^{(n),k}=\dfrac{1}{\nu_{k}}\displaystyle\sum\limits_{t=n-\nu_{k}}^{n-1} {d_{i,j,k}^{(t)}\mathcal{Z}_{k}^{(t)}}  \\
P_{{\rm FA,T},j}^{(n),k}=\dfrac{1}{\nu_{k}}\displaystyle\sum\limits_{t=n-\nu_{k}}^{n-1} {D_{j,k}^{(t)}\mathcal{Z}_{k}^{(t)}}
\end{array},
\begin{array}{l}
   0\le i \le m_{j,k}\\
   j \in\mathbb{M} \\
   k \in \mathbb{B}
 \end{array},
\end{split}
\end{equation}
where $\mathcal{Z}_{j}^{(t)}$  represents the factual status of ${\rm CH}_{k}$ at ${\rm QP}^{(t)}$, whereof the BS has no deterministic knowledge but only estimates it by $D_{j,k}^{(t)}$ using the database readings $R_{j,k}^{(t)}$ and local sensing results.  To have more smooth measures,  the latest $\nu_{k}$  samples derived from ${\rm CH}_{k}$ are averaged in \eqref{Equation14}.
\subsubsection{Misdetection Rate}
Given its probabilistic expression in (\ref{Equation2}), the misdetection rate is experimentally estimated for an individual CPE as well as for the entire cell using the following formulae,
\begin{equation}\label{Equation15}
\begin{split}
\begin{array}{c}
P_{{\rm MD},i,j}^{(n),k}=\dfrac{1}{\nu_{k}}\displaystyle\sum\limits_{t=n-\nu_{k}}^{n-1} {\left(\neg d_{i,j,k}^{(t)}\right)\mathcal{Z}_{k}^{(t)}} \\
P_{{\rm MD,T},j}^{(n),k}=\dfrac{1}{\nu_{k}}\displaystyle\sum\limits_{t=n-\nu_{k}}^{n-1}
{\left(\neg D_{j,k}^{(t)}\right)\mathcal{Z}_{k}^{(t)}}
\end{array},
\begin{array}{l}
 0\le i \le m_{j,k}\\
 j \in\mathbb{M} \\
 k \in \mathbb{B}
\end{array}.
\end{split}
\end{equation}
\subsubsection{Successful Discovery Rate}
The successful probability of discovery quantifies the receiver's ability in correct estimation of the channel status, and is experimentally calculated as,
\begin{equation}\label{Equation16}
\begin{array}{l}
P_{{\rm SD}, i,j}^{(n),k}=\dfrac{1}{\nu_{k}}\displaystyle\sum\limits_{t=n-\nu_{k}}^{n-1} {d_{i,j,k}^{(t)}\odot \mathcal{Z}_{k}^{(t)}}\\
P_{{\rm SD, T},j}^{(n),k}=\dfrac{1}{\nu_{k}}\displaystyle\sum\limits_{t=n-\nu_{k}}^{n-1} {D_{j,k}^{(t)}\odot \mathcal{Z}_{k}^{(t)}}
\end{array},
\begin{array}{l}
 0\le i \le m_{j,k}\\
 j \in\mathbb{M} \\
 k \in \mathbb{B}
\end{array}.
\end{equation}

By probing \eqref{Equation14}-\eqref{Equation16} in every sensing interval, the WRAN BS remains vigilant as to whether certain thresholds are exceeded or not,\footnote{Since said threshold was not explicitly stated in \cite{IEEE80222}, $[ P_{{\rm MD,T}}^{{\rm max}},P_{{\rm FA,T}}^{{\rm max}}]=[0.1,\, 0.1]$ is chosen for the case study in this section.} whereupon appropriate actions such as channel switching, cell shutdown, etc., may be taken.
\subsubsection{Correlation Point Indicator}
The false-alarm and misdetection rates are intrinsically conflicting measures, where an increase of one inevitably results in a decrease of the other. Thus, when comparing different SFMs using these measures, no conclusive decision can be made about the superiority or the inferiority of one versus the other. This fact (also known as bias-variance trade-off in machine learning and accuracy-efficiency trade-off in decision theory) was illustrated in Fig. \ref{fig:FigureIVa}, where the confined areas underneath the Gaussian PDF for null ($\mathcal{H}_0$) and alternative ($\mathcal{H}_1$) hypotheses signify the paradoxical fact that how the optimal choice of the threshold ${\rm SNR}$, of which we have no knowledge, matters in having the right balance between utilization of the spectrum and radiated interference.

Consequently, the need for a single performance measure that jointly and fairly encompasses $P_{\rm FA}$ and $P_{\rm MD}$ is felt. The correlation matrix $C=[{\rm NWCF}_{j,k}^{(n)}]_{\mathfrak{M} \times \mathfrak{B}}$ serves this purpose where ${\rm NWCF}_{j,k}^{(n)}$ quantifies the amount of statistical similarity between two data streams, namely the factual ($\{\mathcal{Z}_{k}^{(t)}\}_{t=n-\nu_k}^{n-1}$) status of ${\rm CH}_k$ and its estimation by ${\rm WRAN}_j$ ($\{D_{j,k}^{(t)}\}_{t=n-\nu_k}^{n-1}$), over the most recent $\nu_k$ ${\rm QP}$s. To have an scalar indicator, we define a metric called network-wide correlation factor (${\rm NWCF}$) by weighted averaging of the correlation matrix over its rows and columns.
\subsubsection{Goodness of Fit}
The next statistical test used to check the accuracy of MC-LDS is the \textit{Pearson} chi-square test for the goodness-of-fit. As one of the most popular and prevalent statistical tests \cite{Chernoff1954}, \textit{Pearson} test is used to verify whether an obtained histogram (or an estimated categorical data) matches a theoretical PDF (or factual data). Represented by $\left[\chi_{j,k}^{(n)}\right]^2$ and called Pearson's cumulative test statistic, a real number is obtained per cell, operating channel, and sensing interval, which can either be used directly to quantify the similarity level according to,
\begin{equation}\label{Equation17}
{\left[\chi_{j,k}^{(n)}\right]}^2=\sum\limits_{t=n-\nu_{k}}^{n-1} \frac{\left( D_{j,k}^{(t)}-\mathcal{Z}_{k}^{(t)} \right)^{2}}{\mathcal{Z}_{k}^{(t)}} , \hspace{1cm}
\begin{array}{l}
 j \in\mathbb{M} \\
 k \in \mathbb{B}
\end{array}
\end{equation}
or in conjunction with a \textit{p-value} \cite{Chernoff1954} to reject/accept whether $D_{j,k}^{(t)}$ is a good enough estimator for $\mathcal{Z}_{k}^{(t)}$. The definition of $\nu_k$ was given earlier.

Given the two data streams, i.e. the factual channel status $\left[ \mathcal{Z}_{k} \right]$ and the estimated channel status $\left[ D_{j,k} \right]$, and considering that the network under study is composed of $\mathfrak{M}$ coexisting cells and $\mathfrak{B}$ channels, a single performance vector  $[{\rm NWCF},P_{{\rm SD,T}},P_{{\rm MD,T}},P_{{\rm FA,T}}, {\chi}^2]$ of five elements is obtained per cell, per channel. Thus, a $\mathfrak{M}\times \mathfrak{B}$ matrix of above vectors is obtained to evaluate the performance of MC-LDS. An example of this matrix for the misdetection rate is extracted from the simulator and reproduced in Table \ref{tab3}.

\begin{table*}[ht]
\begin{center}
\begin{small}
\caption{An instance of misdetection rate matrix $[P_{{\rm MD,T},j}^{(n),k}]_{k \in \mathbb{B},j\in\mathbb{M}}$ for the network in Fig. \ref{fig:FigureI}. The appearance of ${\rm NA}$ at $[j,k]^{\rm th}$ entry indicates that during the simulator runtime $k\notin {\rm OCL}_j$, i.e. no conclusion can be made about ${\rm CH}_{k}$ in ${\rm WRAN}_{j}$.}
\begin{tabular}{|>{\columncolor[gray]{0.8}}l|l|l|l|l|l|l|l|l|l|l|}
\hline
\rowcolor[gray]{0.8}
\cellcolor{white}&
\textbf{CH}$_{{1}}$ &
\textbf{CH}$_{{\, 2}}$ &
\textbf{CH}$_{{\, 3}}$ &
\textbf{CH}$_{{\, 4}}$ &
\textbf{CH}$_{{\, 5}}$ &
\textbf{CH}$_{{\, 6}}$ &
\textbf{CH}$_{{\, 7}}$ &
\textbf{CH}$_{{8}}$ &
\textbf{CH}$_{{9}}$ &
\textbf{CH}$_{{10}}$ \\
\hline
\textbf{WRAN}$_{{1}}$ &
\cellcolor{BrickRed} {\color{white}{\textbf{NA}}}&
0.0470&
\cellcolor{BrickRed} {\color{white}{\textbf{NA}}}&
0.173&
0.225&
0.071&
0.052&
0.038&
0.16&
\cellcolor{BrickRed} {\color{white}{\textbf{NA}}} \\
\hline
\textbf{WRAN}$_{{2}}$ &
0.141&
0.321&
\cellcolor{BrickRed} {\color{white}{\textbf{NA}}}&
\cellcolor{BrickRed} {\color{white}{\textbf{NA}}}&
0.02&
0.165&
0.043&
0.0.72&
0.123&
0.022 \\
\hline
\textbf{WRAN}$_{{3}}$ &
0.0083&
0.258&
0.0033&
0.059&
0.022&
0.029&
\cellcolor{BrickRed} {\color{white}{\textbf{NA}}}&
0.0089&
0.081&
0.081 \\
\hline
\textbf{WRAN}$_{{4}}$ &
0.25&
0.0037&
0.045&
0.24&
0.22&
0.076&
0.016&
0.37&
0.083&
0.068 \\
\hline
\textbf{WRAN}$_{{5}}$ &
0.016&
0.049&
0.049&
\cellcolor{BrickRed} {\color{white}{\textbf{NA}}}&
0.076&
0.31&
0.090&
0.053&
0.093&
0.062 \\
\hline
\textbf{WRAN}$_{{6}}$ &
0.066&
0.29&
0.11&
0.40&
0.083&
0.058&
0.27&
0.29&
0.085&
0.25 \\
\hline
\textbf{WRAN}$_{{7}}$ &
0.065&
\cellcolor{BrickRed} {\color{white}{\textbf{NA}}}&
0.12&
0.20&
0.096&
0.077&
0.34&
\cellcolor{BrickRed} {\color{white}{\textbf{NA}}}&
0.054&
0.060 \\
\hline
\textbf{WRAN}$_{{8}}$ &
0.086&
0.34&
0.076&
0.071&
0.089&
0.17&
0.060&
0.079&
0.055&
\cellcolor{BrickRed} {\color{white}{\textbf{NA}}} \\
\hline
\textbf{WRAN}$_{{9}}$ &
0.089&
0.073&
0.39&
0.060&
0.072&
0.090&
0.11&
0.17&
0.26&
0.40 \\
\hline
\textbf{WRAN}$_{{10}}$ &
\cellcolor{BrickRed} {\color{white}{\textbf{NA}}}&
0.28&
0.051&
0.085&
0.20&
0.080&
0.31&
0.12&
0.13&
0.26 \\
\hline
\textbf{WRAN}$_{{11}}$&
0.098&
0.084&
0.073&
0.057&
0.22&
0.37&
0.072&
0.071&
0.081&
0.12 \\
\hline
\textbf{WRAN}$_{{12}}$ &
\cellcolor{BrickRed} {\color{white}{\textbf{NA}}}&
0.32&
0.086&
0.19&
0.11&
0.065&
\cellcolor{BrickRed} {\color{white}{\textbf{NA}}}&
0.10&
0.052&
0.094 \\
\hline
\end{tabular}
\label{tab3}
\end{small}
\end{center}
\end{table*}

\begin{figure}[t!]
\centering
\subfloat[Network wide correlation factor]{\includegraphics[scale=0.45]{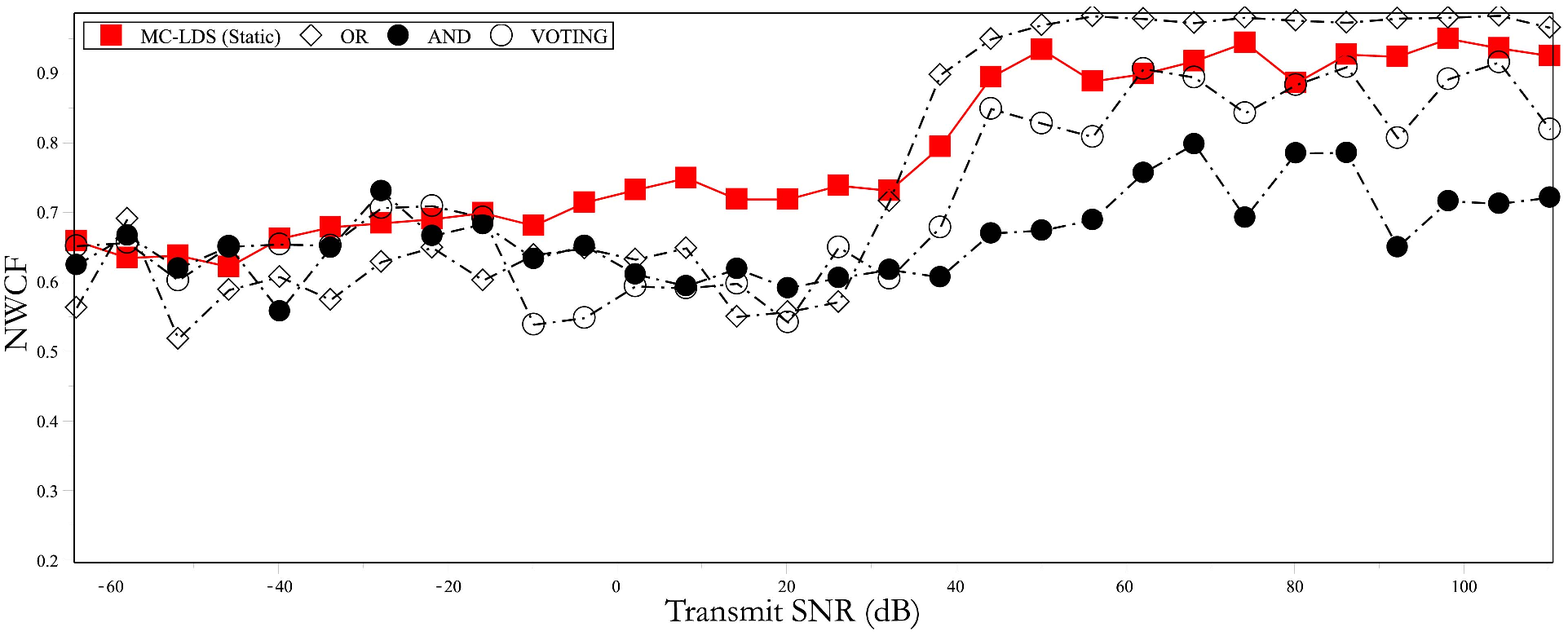} \label{fig:FigureVIIa}}\\ \vspace{-10 pt}
\subfloat[Network wide successful discovery rate $P_{\rm SD}$]{\includegraphics[scale=0.45]{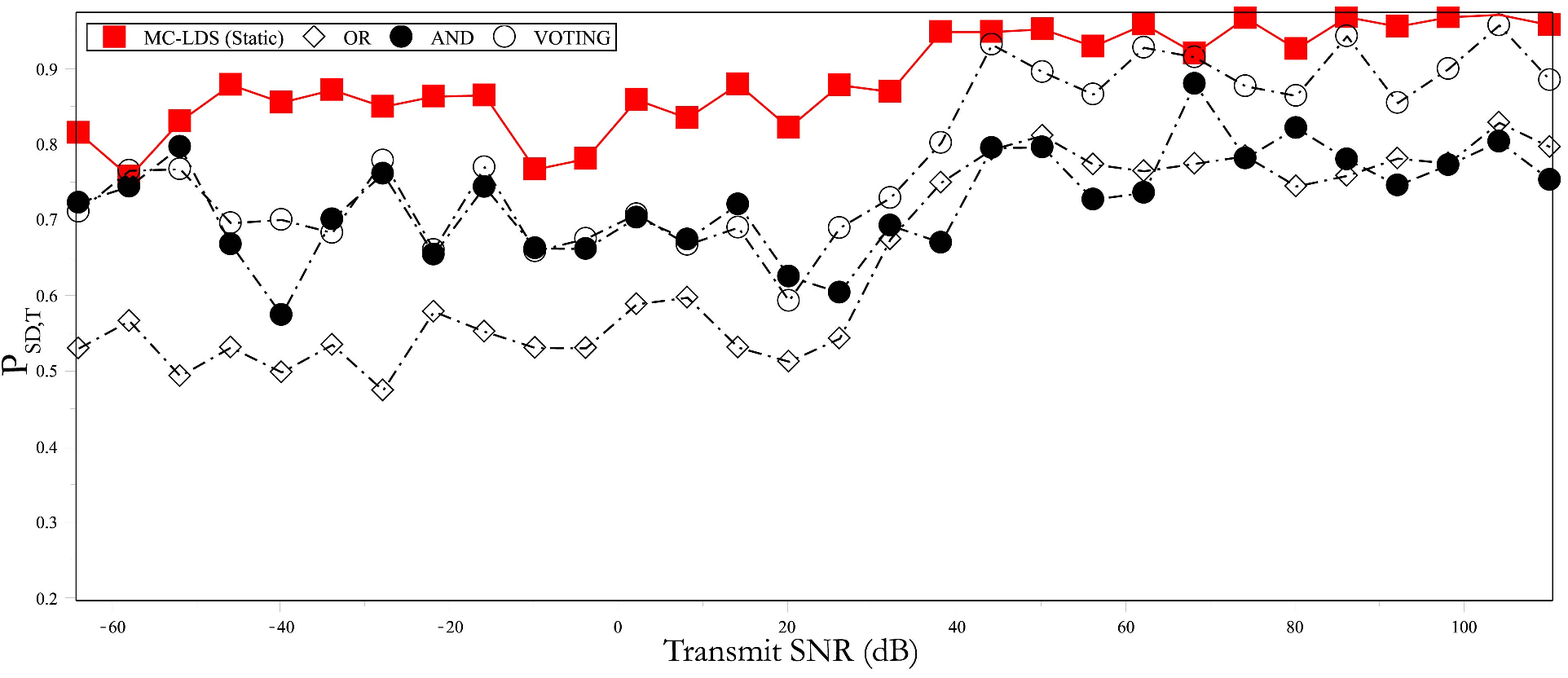} \label{fig:FigureVIIb}} \\ \vspace{-10 pt}
\subfloat[Network wide misdetection rate $P_{\rm MD}$]{\includegraphics[scale=0.45]{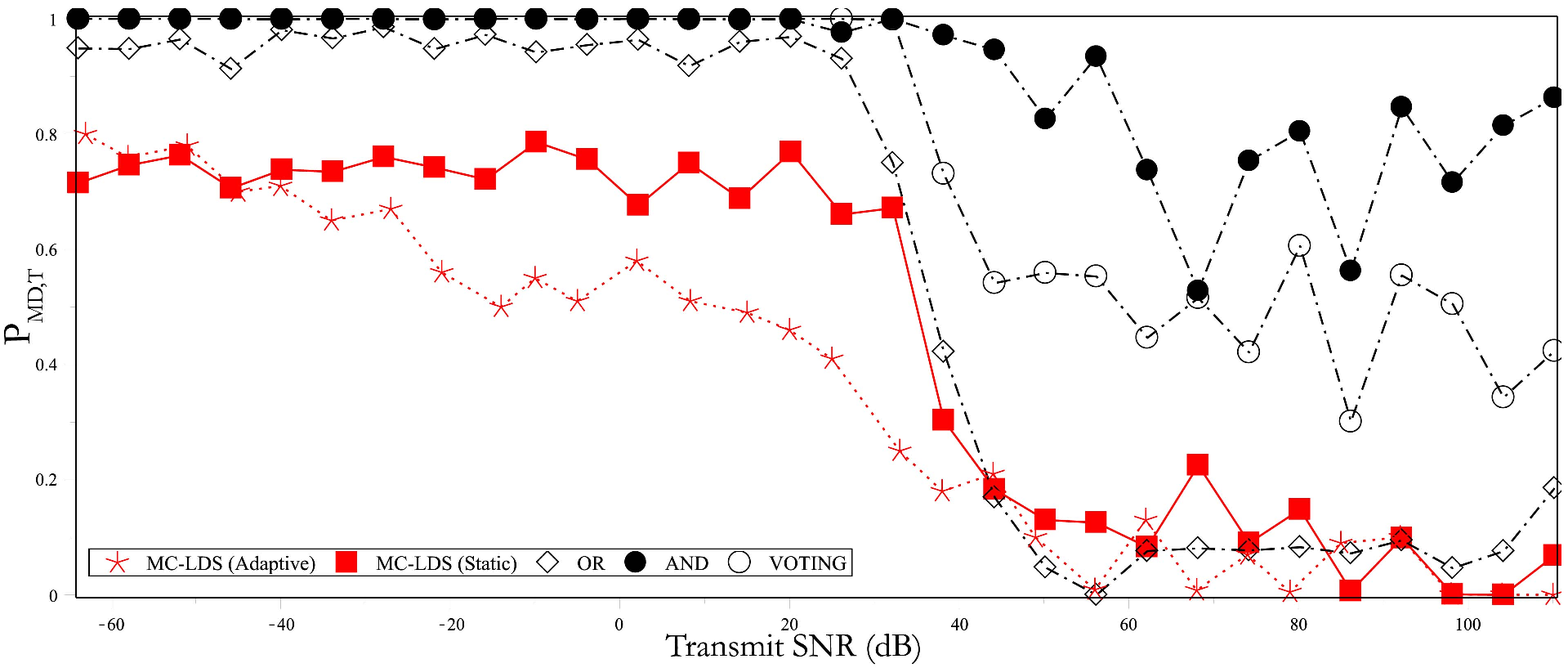} \label{fig:FigureVIIc}} \\ \vspace{-10 pt}
\subfloat[Network wide false-alarm rate $P_{\rm FA}$]{\includegraphics[scale=0.45]{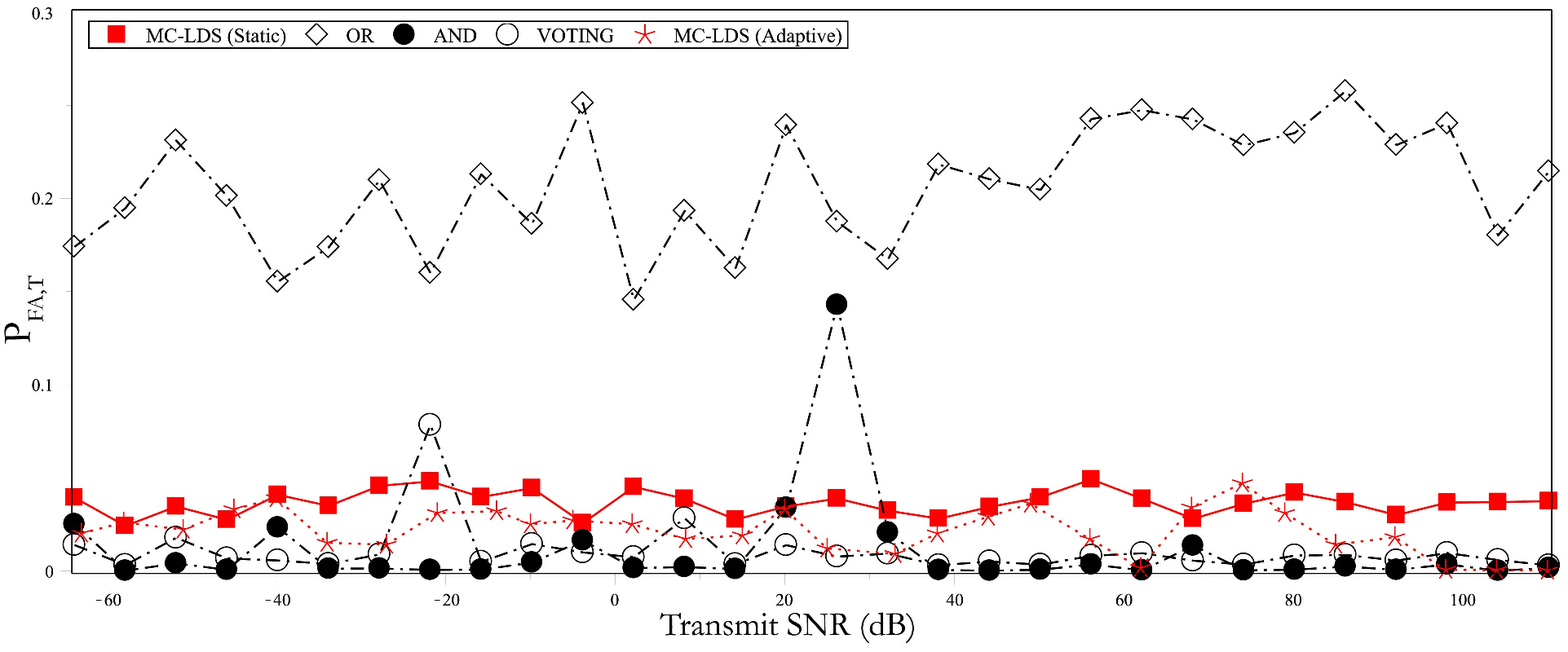} \label{fig:FigureVIId}} \\\vspace{-10 pt}
\subfloat[Goodness of Fit]{\includegraphics[scale=0.45]{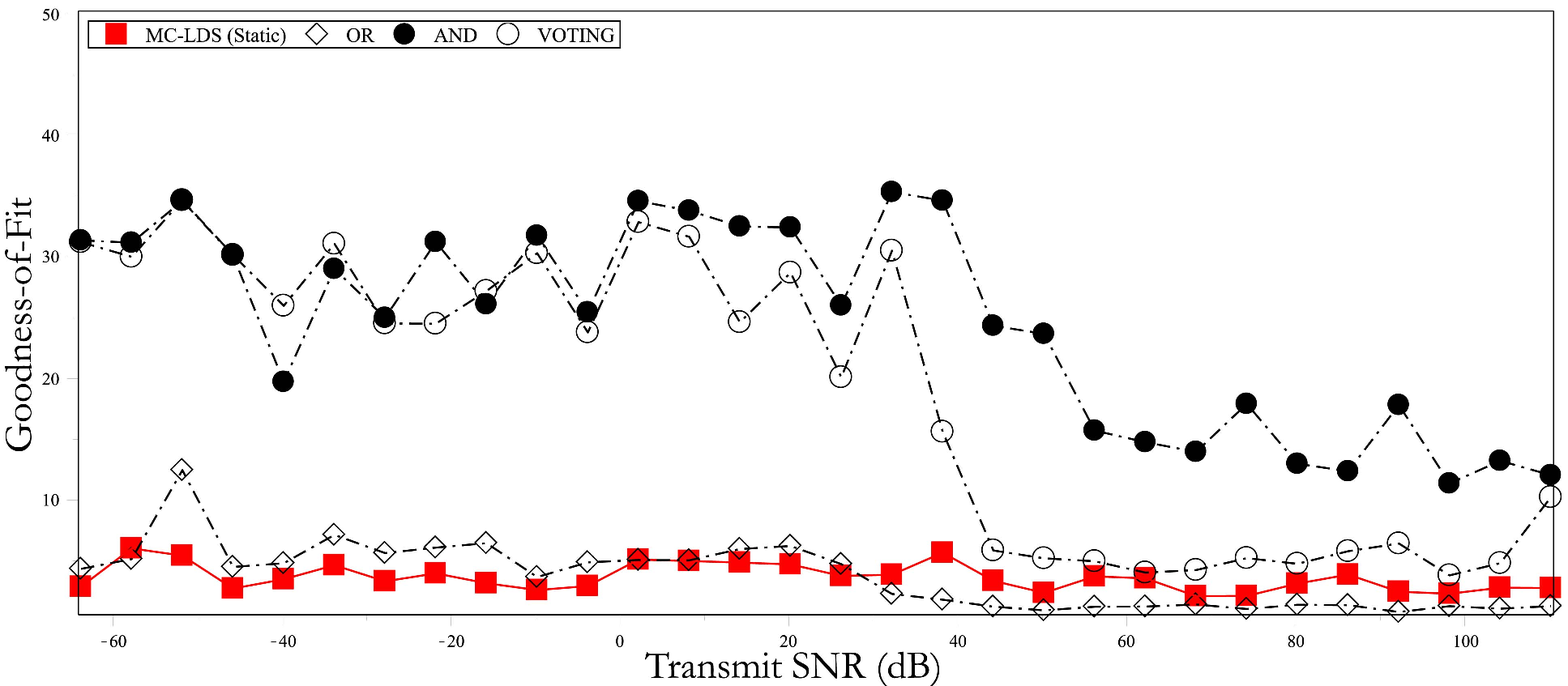} \label{fig:FigureVIIe}}
\caption{Performance comparison of AND, OR, VOTING and MC-LDS SMFs by different performance metrics.}
\label{fig:FigureVII}
\end{figure}
\subsection{Comparative Results}
Numerous experiments were conducted to verify the performance of MC-LDS. Comparative results were plotted versus ${\rm SNR}$ at transmitter side, often known as transmit ${\rm SNR}$ (dB). For the sake of comprehensiveness, different circumstances were examined, including diverse ${\rm SNR}$ regimes, fading situations, incumbent activity pattern profiles, etc. Furthermore, the performance of MC-LDS was probed in the presence of hidden and exposed terminals. The observations unanimously prove that MC-LDS is superior to AND, OR and VOTING SFMs. Of particular importance is the performance uniformity of MC-LDS in different ${\rm SNR}$ regimes.

Fig. \ref{fig:FigureVII} proves the above claim for a wide ${\rm SNR}$ range ($-70$ to $110$ dB). Starting from low ${\rm SNR}$ regimes, wherein, principally, the real merits or weaknesses of any decision combining method lie, the MC-LDS SFM is unequivocally superior to all its rivals for all the metrics. At high ${\rm SNR}$s, MC-LDS still keeps the superiority in comparison to VOTING and AND rules, even though, correlation-wise, the OR-rule has the edge over MC-LDS.  Notwithstanding, this slight superiority is severely compromised by the unacceptably high false-alarm rate that OR-rule exhibits at all the times. This is further to the fact that when ${\rm IAR}<<1$ and the incumbent network's alteration frequency (${\rm IAF}$) is high,\footnote{The incumbent alteration frequency is the frequency with which ON and OFF periods alternate, whereas the incumbent activity rate indicates how relatively large ON periods are w.r.t. OFF periods.} the OR-rule exhibits extremely poor performance in all the metrics while MC-LDS still keeps the edge over the others. Overall, despite the fact that achieving low false-alarm and misdetection rates at the same time is theoretically impossible, as discussed earlier, the MC-LDS mechanism still finds the best trade-off that maximizes the ${\rm NWCF}$ and minimizes the {\it Pearson}'s goodness-of-fit metric.

One should note that the superior behavior of the OR-rule in terms of misdetection rate does not make it an optimal decision combining approach due to the very high false-alarm rate it manifests, as shown in Fig. \ref{fig:FigureVIIc}, \ref{fig:FigureVIId}. The same is true about the AND-rule which yields low false-alarm rate at the expense of unacceptably high misdetection rate. More troublesome is the unstable behavior of these two SFMs when ${\rm IAR}$ or ${\rm IAF}$ vary, exhibiting  very different performance profiles. Yet, one should keep in mind that  the joint inclusion of OR/AND in comparisons is always a good benchmark as it reveals the flaws and strengths of any decision combining mechanism, including MC-LDS, a fact that was proven in \cite{Visotsky2005} and exploited in many papers, e.g. \cite{Peh2010}. For the sake of confidence, the comparisons were also made against the VOTING-rule, which was shown in \cite[P.63]{Varshney1996}, \cite[Ch.5]{Biglieri2013} and\cite{Lunden2007} to secure sub-optimal outcomes in many situations.
\subsection{Adaptive Tuning}
The two degrees of freedom, the temporal discount factor $\alpha$ and the historic count $N$, can be chosen as control variables to give MC-LDS a great deal of flexibility. Our investigations confirm that in situations where the achieved performance vector is not satisfactory, for instance when there is no reasonable balance between $P_{{\rm MD},T}$ and $P_{{\rm FA},T}$, the dynamic adjustment of $\alpha$ and $N$ can make drastic changes. One approach  is to associate these parameters with the estimated false-alarm and misdetection rates in a linear fashion, such that
\begin{equation}\label{Equation18}
\begin{split}
\begin{array}{l}
N_{j,k}=\left\lfloor b-a\cdot \overline{P_{{\rm MD,T},j}^{(n),k}}\right\rfloor\\\\
\alpha_{j,k}=c+d\cdot \overline{P_{{\rm FA,T},j}^{(n),k}}
\end{array},
\hspace{1cm}
\begin{array}{l}
 0\le i \le m_{j,k}\\
 j \in\mathbb{M} \\
 k \in \mathbb{B}
\end{array},
\end{split}
\end{equation}
where $\lfloor~\rfloor$ is the \textit{floor} operation, $b-a>1$, $c+d<1$ and $a, b, c, d \in \mathbb{R}^{+}$  are chosen statically. Red dotted curves with star marks in Fig. \ref{fig:FigureVIIc} and \ref{fig:FigureVIId} illustrate the obtainable gain, in terms of $P_{\rm FA,T}$ and $P_{\rm MD,T}$, by simply feeding the output of the system back to its input, and substantiate the importance of feedback in stabilizing cognitive systems and improving their performance.

\section{Conclusion}
This work proposed an efficient decision combining method, named multi-channel learning-based distributed sensing (MC-LDS) mechanism, for wireless regional area networks (WRANs). While conforming with the IEEE 802.22 standard directives, MC-LDS operates based on a simple learning concept that uses differentiation as well as reward-penalty procedures to intelligently combine current and past sensing measurements. The combination was done in such a way that both accurate and faulty sensors contribute constructively toward an accurate final decision. To substantiate the merits of MC-LDS, a WRAN simulator was built. Through precise probing of a comprehensive set of quantities such as false-alarm, misdetection and successful discovery rates, as well as correlation factor and goodness-of-fit metric, it was demonstrated that MC-LDS is considerably superior to AND, OR and VOTING rules. Among the advantages of MC-LDS,  its accuracy, implementation ease, fairness, adjustability and stability are to be emphasized. Having shown the intrinsic inefficiencies and drawback of the SFMs suggested by the IEEE 802.22 standard, we present MC-LDS as a competent candidate for adoption in WRANs. Furthermore, MC-LDS can be integrated to boost the sensing performance in other promising technologies and standards such as \textit{White-Fi} (IEEE 802.11af WLANs) wireless personal area networks and ZigBee (IEEE 802.15 family), cognitive WiMax (IEEE 802.16h) and the recent IEEE 1900.6b standard emerged to support spectrum databases using spectrum sensing information.

\bibliographystyle{IEEEtran}


\begin{biography}
[{\includegraphics[width=1in,height=1.25in,clip,keepaspectratio]{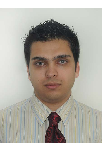}}]
{Navid Tadayon} (S'10) received his B.Sc. degree in electrical engineering, Telecommunications, from Ferdowsi University, Mashhad, Iran, in 2006, and his M.Sc. degree from University of Massachusetts Dartmouth, USA, in 2011. He is now working toward his Ph.D. at the Institut National de la Recherche Scientifique-{\it Energy, Materials and Telecommunications} (INRS-EMT), University of Quebec, Montreal, QC, Canada. From 2008 to 2010, he was a Researcher with the Iran Telecommunication Research Center (ITRC). He is currently an associate researcher at \'{E}cole de Technologie Sup\'{e}rieure. His research interests include modeling and analysis of wireless networks as well as designing mechanisms and algorithms for these networks, with a focus on 5G enabling technologies such as cognitive radio networks, HetNets and D2D.

He is recipient of several prestigious Canadian awards, including NSERC post-doctoral fellowship and FRQNT Ph.D. Merit scholarship. The outcomes of his investigations have been a book and dozens of papers published in distinguished journals and flagship conferences.
\end{biography}

\begin{biography}[{\includegraphics[width=1in,height=1.25in,clip,keepaspectratio]{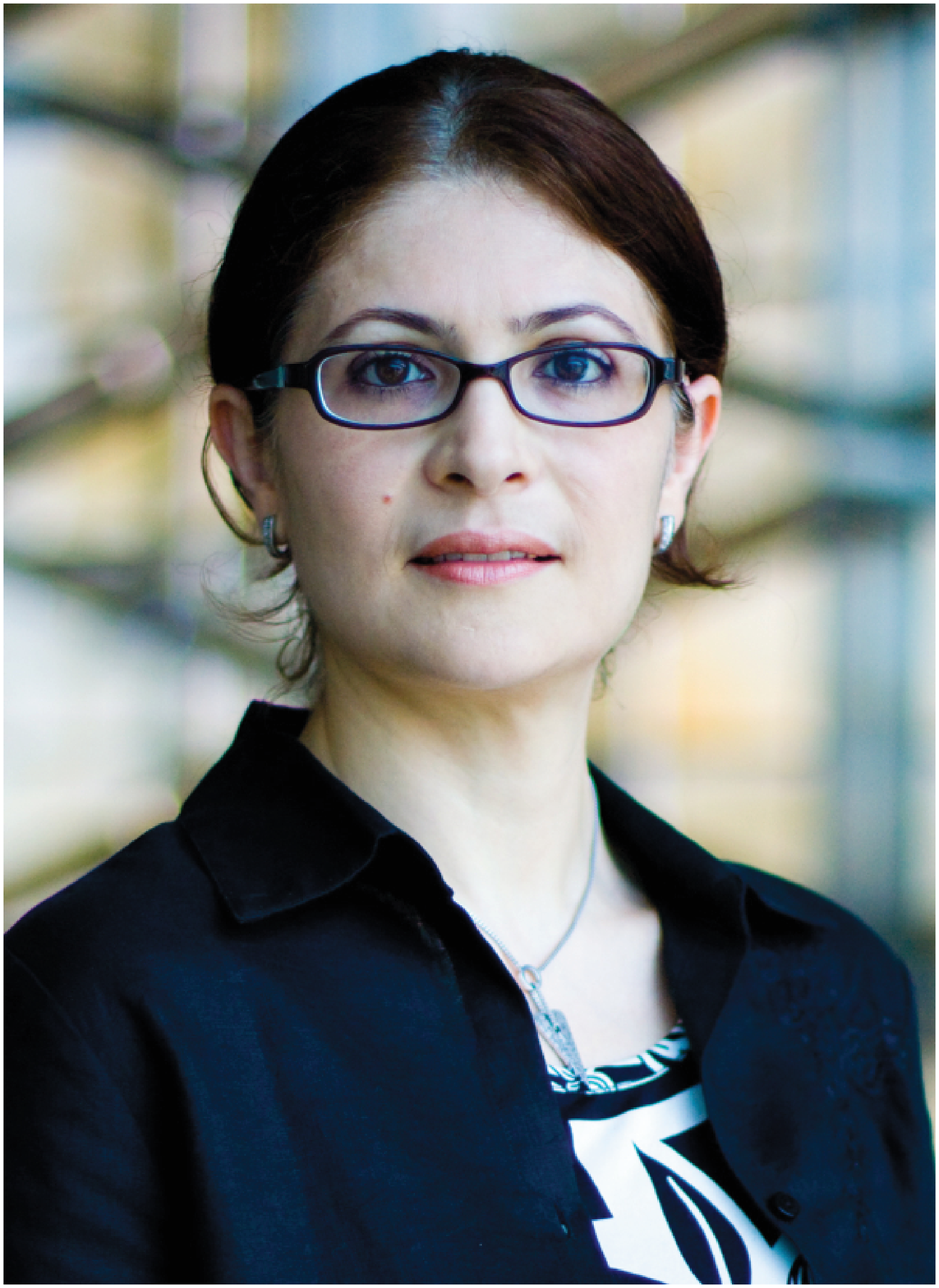}}]
{Sonia A\"{\i}ssa} (S'93-M'00-SM'03) received her Ph.D. degree in Electrical and Computer Engineering from McGill University, Montreal, QC, Canada, in 1998. Since then, she has been with the Institut National de la Recherche Scientifique-{\it Energy, Materials and Telecommunications} Center (INRS-EMT), University of Quebec, Montreal, QC, Canada, where she is a Full Professor.

From 1996 to 1997, she was a Researcher with the Department of Electronics and Communications of Kyoto University,
and with the Wireless Systems Laboratories of NTT, Japan. From 1998 to 2000, she was a Research Associate at INRS-EMT. In 2000-2002, while she was an Assistant Professor, she was a Principal Investigator in the major program of personal and mobile communications of the Canadian Institute for Telecommunications Research, leading research in radio resource management for wireless networks. From 2004 to 2007, she was an Adjunct Professor with Concordia University, Montreal. She was Visiting Invited Professor at Kyoto University, Japan, in 2006, and Universiti Sains Malaysia, in 2015.
Her research interests include the modeling, design and performance analysis of wireless communication systems and networks.

Dr. A\"{\i}ssa is the Founding Chair of the IEEE Women in Engineering Affinity Group in Montreal, 2004-2007; acted as TPC Symposium Chair or Cochair at IEEE ICC '06 '09 '11 '12; Program Cochair at IEEE WCNC 2007; TPC Cochair of IEEE VTC-spring 2013; and TPC Symposia Chair of IEEE Globecom 2014. Her main editorial activities include: Editor, {\scshape IEEE Transactions on Wireless Communications}, 2004-2012; Associate Editor and Technical Editor, {\scshape IEEE Communications Magazine}, 2004-2015; Technical Editor, {\scshape IEEE Wireless Communications Magazine}, 2006-2010; and Associate Editor, {\it Wiley Security and Communication Networks Journal}, 2007-2012. She currently serves as Area Editor for the {\scshape IEEE Transactions on Wireless Communications}. Awards to her credit include the NSERC University Faculty Award in 1999; the Quebec Government FRQNT Strategic Faculty Fellowship in 2001-2006; the INRS-EMT Performance Award multiple times since 2004, for outstanding achievements in research, teaching and service; and the Technical Community Service Award from the FQRNT Centre for Advanced Systems and Technologies in Communications, 2007. She is co-recipient of five IEEE Best Paper Awards and of the 2012 IEICE Best Paper Award; and recipient of NSERC Discovery Accelerator Supplement Award. She is a Distinguished Lecturer of the IEEE Communications Society (ComSoc) and an Elected Member of the ComSoc Board of Governors. Professor A\"{\i}ssa is a Fellow of the Canadian Academy of Engineering.
\end{biography}

\end{document}